\documentclass{aa}

\usepackage{txfonts}

\usepackage{hyperref}
\hypersetup{
    colorlinks=true,
    linkcolor=blue,
    citecolor=blue,
    urlcolor=blue,
    }
    
\usepackage[T1]{fontenc}

\DeclareRobustCommand{\VAN}[3]{#2}
\let\VANthebibliography\thebibliography
\def\thebibliography{\DeclareRobustCommand{\VAN}[3]{##3}\VANthebibliography}

\usepackage{graphicx}	% Including figure files
\usepackage{amsmath}	% Advanced maths commands
\usepackage{amssymb}	% Extra maths symbols
\usepackage{xcolor}

\newcommand{\nii}{[N\,{\sc{ii}}]}
\newcommand{\oiii}{[O\,{\sc{iii}}]}
\newcommand{\ha}{H$\alpha$}
\newcommand{\hb}{H$\beta$}
\newcommand{\siid}{[S\,{\sc{ii}}]$\lambda\lambda6717,6731$}
\newcommand{\sii}{[S\,{\sc{ii}}]}
\newcommand{\heii}{He\,{\sc{ii}}$\lambda4686$}
\newcommand{\hei}{He\,{\sc{i}}}
\newcommand{\ariv}{[Ar\,{\sc{iv}}]}
\newcommand{\micron}{\mu m}

\begin{document}
%%%%%%%%%%%%%%%%%%% TITLE PAGE %%%%%%%%%%%%%%%%%%%

\title{GA-NIFS: A massive black hole in a low-metallicity AGN at $z\sim5.55$ revealed by {\it JWST}/NIRSpec IFS}

   \author{Hannah \"Ubler\inst{\ref{Kavli},\ref{Cavendish}}\thanks{E-mail: hu215@cam.ac.uk} \and
   Roberto Maiolino\inst{\ref{Kavli},\ref{Cavendish},\ref{RM3}} \and
   Emma Curtis-Lake\inst{\ref{ECL}} \and
   Pablo G.~P\'erez-Gonz\'alez\inst{\ref{SA}} \and
   Mirko Curti\inst{\ref{MC},\ref{Kavli},\ref{Cavendish}} \and
   Michele Perna\inst{\ref{SA}} \and
   Santiago Arribas\inst{\ref{SA}} \and
   Stéphane Charlot\inst{\ref{SC}} \and
   Madeline A.~Marshall\inst{\ref{MM1},\ref{MM2}} \and
   Francesco D'Eugenio\inst{\ref{Kavli},\ref{Cavendish}} \and   
   Jan Scholtz\inst{\ref{Kavli},\ref{Cavendish}} \and
   Andrew Bunker\inst{\ref{AB}} \and
   Stefano Carniani\inst{\ref{pisa}} \and   
   Pierre Ferruit\inst{\ref{PF}} \and
   Peter Jakobsen\inst{\ref{PJ1},\ref{PJ2}} \and
   Hans-Walter Rix\inst{\ref{MPIA}}\and 
   Bruno Rodr\'iguez Del Pino\inst{\ref{SA}} \and  
   Chris J.~Willott\inst{\ref{MM1}} \and  
   Torsten B\"{o}ker\inst{\ref{ESA}} \and
   Giovanni Cresci\inst{\ref{INAF}} \and
   Gareth C. Jones\inst{\ref{AB}} \and
   Nimisha Kumari\inst{\ref{NK}} \and
   Tim Rawle\inst{\ref{ESA}}
}

\institute{Kavli Institute for Cosmology, University of Cambridge, Madingley Road, Cambridge, CB3 0HA, UK\label{Kavli} \and
Cavendish Laboratory, University of Cambridge, 19 JJ Thomson Avenue, Cambridge, CB3 0HE, UK\label{Cavendish} \and
Department of Physics and Astronomy, University College London, Gower Street, London WC1E 6BT, UK\label{RM3} \and
Centre for Astrophysics Research, Department of Physics, Astronomy and Mathematics, University of Hertfordshire, Hatfield, AL10 9AB, UK\label{ECL} \and
Centro de Astrobiolog\'{\i}a (CAB), CSIC-INTA, Ctra. de Ajalvir km 4, Torrej\'on de Ardoz, E-28850, Madrid, Spain\label{SA} \and
European Southern Observatory, Karl-Schwarzschild-Straße 2, 85748, Garching, Germany\label{MC} \and
Sorbonne Universit\'e, CNRS, UMR 7095, Institut d’Astrophysique de Paris, 98 bis bd Arago, 75014 Paris, France\label{SC} \and
National Research Council of Canada, Herzberg Astronomy \& Astrophysics Research Centre, 5071 West Saanich Road, Victoria, BC V9E 2E7, Canada\label{MM1} \and
ARC Centre of Excellence for All Sky Astrophysics in 3 Dimensions (ASTRO 3D), Australia\label{MM2} \and
University of Oxford, Department of Physics, Denys Wilkinson Building, Keble Road, Oxford OX13RH, UK\label{AB} \and
Scuola Normale Superiore, Piazza dei Cavalieri 7, I-56126 Pisa, Italy\label{pisa} \and
European Space Agency, ESAC, Villanueva de la Ca\~{n}ada, E-28692 Madrid, Spain\label{PF} \and
Cosmic Dawn Center (DAWN), Copenhagen, Denmark \label{PJ1} \and Niels Bohr Institute, University of Copenhagen, Jagtvej 128, DK-2200, Copenhagen, Denmark \label{PJ2} \and
Max-Planck-Institut f\"{u}r Astronomie, K\"onigstuhl 17, 69117, Heidelberg, Germany\label{MPIA} \and
European Space Agency, c/o STScI, 3700 San Martin Drive, Baltimore, MD 21218, USA\label{ESA} \and
INAF - Osservatorio Astrofisco di Arcetri, largo E. Fermi 5, 50127 Firenze, Italy\label{INAF} \and
AURA for the European Space Agency, Space Telescope Science Institute, Baltimore, Maryland, USA\label{NK} 
}

\abstract{We present {\it JWST}/NIRSpec Integral Field Spectrograph rest-frame optical data of the compact $z=5.55$ galaxy GS\_3073. Its prominent broad components in several hydrogen and helium lines (while absent in the forbidden lines), and the detection of a large equivalent width of He\,{\sc{ii}}$\lambda4686$, EW(He\,{\sc{ii}}) $\sim20$\AA, unambiguously identify it as an active galactic nucleus (AGN).
We measure a gas-phase metallicity of $Z_{\rm gas}/Z_\odot\sim0.21^{+0.08}_{-0.04}$, lower than what has been inferred for both more luminous AGN at similar redshift and lower redshift AGN.
We empirically show that classical emission line ratio diagnostic diagrams cannot be used to distinguish between the primary ionisation source (AGN or star formation) for such low-metallicity systems, whereas different diagnostic diagrams involving He\,{\sc{ii}}$\lambda4686$ prove very useful, independent of metallicity.
We measure the central black hole mass to be $\log(M_{\rm BH}/M_\odot)\sim8.2\pm0.4$ based on the luminosity and width of the broad line region of the H$\alpha$ emission. While this places GS\_3073 at the lower end of known high-redshift black hole masses, it still appears to be over-massive compared to its host galaxy properties.
We detect an outflow with projected velocity $\gtrsim700$~km/s and infer an ionised gas mass outflow rate of about $100\ M_\odot/$yr, suggesting that GS\_3073 is able to enrich the intergalactic medium with metals one billion years after the Big Bang.}
  

\maketitle

%%%%%%%%%%%%%%%%%%%%%%%%%%%%%%%%%%%%%%%%%%%%%%%%%%

%%%%%%%%%%%%%%%%% BODY OF PAPER %%%%%%%%%%%%%%%%%%

\section{Introduction}\label{s:intro}

Emission line ratio diagnostics are a key tool to measure physical conditions in the interstellar medium (ISM). At $z=0$, this has led to the development of rest-frame optical diagnostics such as \oiii$\lambda5007$/\hb\ {\it vs.} \nii$\lambda6584$/\ha\ \citep[the so-called BPT diagram;][]{Baldwin81}, \oiii$\lambda5007$/\hb\ {\it vs.} \siid/\ha\ and \oiii$\lambda5007$/\hb\ {\it vs.} [O\,{\sc{i}}]$\lambda6300$/\ha\ \citep{Veilleux87}, which are commonly used to identify whether the primary ionisation source in the ISM is from star formation (SF), photoionization from an active galactic nucleus (AGN), or possibly other ionization sources such as shocks and evolved stars. 
Through multi-object spectrographs such as KMOS \citep{Sharples04, Sharples13} or MOSFIRE \citep{McLean10, McLean12}, the extension of these classifications up to $z\sim3$ became possible \citep[e.g.][]{Kewley13b, Steidel14, Strom17, Curti20, Runco22}.

At $z\sim0$, the hard ionising radiation from AGN accretion discs (generally accompanied by a high ionisation parameter $U$) increases both \nii/\ha\ and \oiii/\hb\ ratios with respect to galaxies dominated by star formation \citep[e.g.][]{Kewley01, Kauffmann03}. 
However, with increasing redshift, also star-forming galaxies (SFGs) tend to move to higher values of \oiii/\hb\ and/or \nii/\ha, possibly as a consequence of $\alpha$-enhanced stellar populations which result in a harder radiation field, variations in the nitrogen abundance, higher electron densities, or a higher ionisation parameter influenced by the SF history \citep[e.g.][]{Brinchmann08a, Kewley13, Steidel14, Masters16, Hirschmann17, Kashino17, Strom17, Topping20, Curti22, HaydenPawson22, Runco22}.
The situation may change more dramatically at higher redshift due to the steadily decreasing metallicity of the ISM, which can result in SFGs changing their location on the line ratio diagnostic diagrams \citep[e.g.][]{Kewley13, Feltre16, Gutkin16, Hirschmann17, Hirschmann19, Hirschmann22, Nakajima22} as is indeed observed in recent {\it JWST} data at $z\sim6-9$ \citep{Curti23, Sanders23, Cameron23}.

At $z>3$, the rest-frame optical emission line properties of AGN and SFGs and their location on diagnostic diagrams have essentially been unexplored so far, due to the technical difficulties of accessing the relevant redshifted lines at z$>$3. However, models expect that their properties should change strongly, primarily because of the lower metallicity at such early epochs \citep[e.g.][]{Groves06, Feltre16, Hirschmann19, Nakajima22}.
{\it JWST} now enables access to the main rest-frame optical lines up to $z\sim7$, and exploring the location of $z>3$ AGN and SFGs in the classical lines ratio diagnostic diagrams is of great importance for our understanding of the role of ionising sources in early galaxies.

Another area in which JWST is expected to enable major progress is the evolution of the scaling relations between supermassive black holes (SMBHs) and their host galaxies. Past studies have shown that at high redshift ($z\gtrsim6$) SMBHs tend to be over-massive relative to their host galaxies, when compared with local relations, by even more than one order of magnitude \citep[e.g.][]{Bongiorno14, Wang16, Shao17, Venemans17, Decarli18, Izumi21}. 
However, it has been suggested that these results might be due to luminous quasars residing in the tail end of the black hole mass ($M_{\rm BH}$) -- host mass ($M_{\rm galaxy}$) distribution. Indeed, similar studies with lower-luminosity quasars have found $M_{\rm BH}/M_{\rm galaxy}$ ratios more consistent with the local relation \citep{Willott15, Willott17, Izumi18}. {\it JWST} offers the possibility of exploring such relations in even lower luminosity AGN at $3<z<9$ and therefore to test these scenarios.

In this paper, we use data from the {\it JWST}/NIRSpec Integral Field Spectrograph \citep[IFS;][]{Jakobsen22, Boeker22} of the galaxy GS\_3073 at $z=5.55$, within the NIRSpec IFS GTO programme. This source is located in the CANDELS/GOODS-S field \citep{Koekemoer11}, 
and has also been referred to as GDS~J033218.92-275302.7 \citep{Vanzella10}. It showed indications for the presence of an AGN in previous spectroscopic studies of its UV light, through detection of very high ionisation lines such as N\,{\sc{v}}$\lambda1240$ and O\,{\sc{vi}}$\lambda1032,1038$ \citep[see][]{Vanzella10, Grazian20}. However, it was not selected for its potential AGN signatures in our survey, being undetected in deep {\it Chandra} $X-$ray observations (2 Ms observations as discussed by \citealp{Vanzella10}; see also  \citealp{Grazian20} for deeper 7 Ms observations). In addition to confirming its AGN nature, the data shed light on its physical conditions and the power source of its ionised gas emission. 
We describe the target GS\_3073 and our {\it JWST} observations in Section~\ref{s:data}. In Section~\ref{s:analysis} we discuss the emission line features and spectral fitting. We describe our measurements of black hole mass, host galaxy dynamical mass, outflow properties, and physical conditions in Section~\ref{s:measurements}. We discuss our results regarding emission line ratio diagnostic diagrams, early black holes, their feedback and enrichment in Section~\ref{s:discussion}, and we conclude in Section~\ref{s:conclusions}.

Throughout, we adopt a \cite{Chabrier03} initial mass function ($0.1-100 M_\odot$) and a flat $\Lambda$CDM cosmology with $H_0=70$~km~s$^{-1}$~Mpc$^{-1}$, $\Omega_\Lambda=0.7$, and $\Omega_m=0.3$.
The wavelengths of emission lines are quoted in rest-frame, if not specified otherwise.

\begin{figure*}
	\centering
	\includegraphics[width=\textwidth]{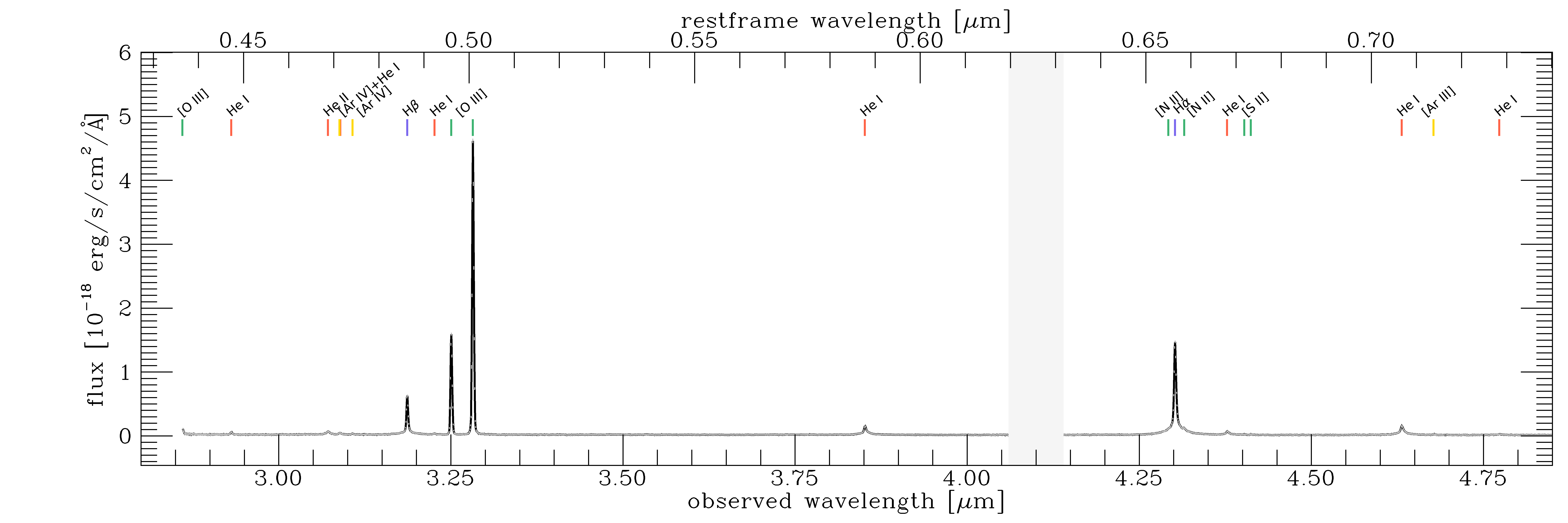}
	\includegraphics[width=\textwidth]{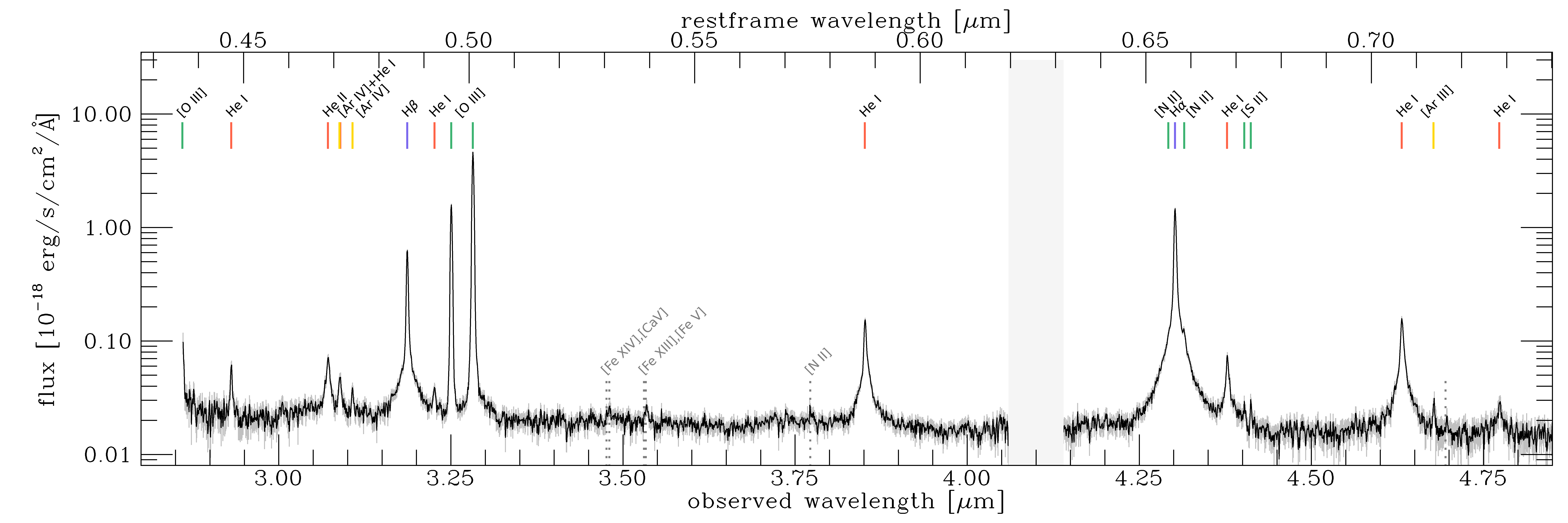}
	\caption{Integrated spectrum extracted from the central three by three spaxels in the wavelength range $2.86\micron < \lambda < 4.85\micron$ with flux in linear scale (top) and log scale (bottom). Several emission lines are present and indicated by vertical lines at the top. We detect seven He\,{\sc{i}} lines, He\,{\sc{ii}}$\lambda4686$, H$\beta$, [O\,{\sc{iii}}]$\lambda\lambda4959,5007$, H$\alpha$, [N\,{\sc{ii}}]$\lambda\lambda6548,6583$, and [S\,{\sc{ii}}]$\lambda\lambda6716,6731$. We also report the detection of [Ar\,{\sc{iv}}]$\lambda4711$, [Ar\,{\sc{iv}}]$\lambda4740$ and [Ar\,{\sc{iii}}]$\lambda7136$ (note that [Ar\,{\sc{iv}}]$\lambda4711$ is blended with  He\,{\sc{i}}$\lambda4713$). The redshifted auroral line [O\,{\sc{iii}}]$\lambda4363$ is only partly covered by the spectral band of our observation with the G395H grating. [O\,{\sc{i}}]$\lambda6003$ falls into the detector gap masked here in the region $4.06\micron < \lambda < 4.14\micron$.  BLR components are present in H$\beta$, H$\alpha$, He\,{\sc{ii}}, and the He\,{\sc{i}} lines. In addition, an outflow component is present, best visible in the broadened, asymmetric line base of the [O\,{\sc{iii}}] doublet. We indicate the positions of possible coronal lines [Fe\,{\sc{xvi}}], [Ca\,{\sc{v}}], [Fe\,{\sc{xiii}}], [Fe\,{\sc{v}}], of the auroral line [N\,{\sc{ii}}]$\lambda5755$, and of another line at $\lambda\sim7167.5$\AA\, the position of which is consistent with Si\,{\sc{i}}, as grey dotted vertical lines in the bottom panel.}
	\label{f:3x3spec}
\end{figure*}

\section{Data}\label{s:data}

\subsection{Ancillary data and previous studies}\label{s:ancillary}

Multi-wavelength observations of GS\_3073 (R.A.\ $3^{\rm h}32^{\rm m}18.93^{\rm s}$, Dec.\ $-27^{\circ}53^{\prime}2.96\arcsec$) are available thanks to the wide imaging coverage of the GOODS-South field, and spectral energy distribution (SED) fitting including photometry from $U$ to radio bands has been performed by several groups \citep[e.g.][]{Stark07, Wiklind08, Raiter10, Vanzella10, Faisst20, Barchiesi22}. While initial results favoured a massive galaxy with an old stellar population, later work reported evidence for two stellar populations, including a younger one, with a total stellar mass of $\log(M_\star/M_\odot)\sim10.5-10.7$, and a star formation rate of SFR~$\sim30-120 M_\odot/$yr. With our new data, we will show that these estimates require some revision.

Spectroscopic observations in the rest-frame UV with FORS2 and VIMOS \citep{LeFevre03} have found evidence for a galactic wind traced by Ly$\alpha$, and for the presence of several high-ionisation species such as O\,{\sc{vi}}$\lambda1032$, N\,{\sc{v}}$\lambda1240$, and N\,{\sc{iv}}]$\lambda\lambda1483,1486$ \citep{Raiter10, Vanzella10, Grazian20, Barchiesi22}.
In addition, GS\_3073 has been detected in [C\,{\sc{ii}}]$158\micron$ without continuum detection within the ALPINE survey \citep{LeFevre20, Faisst20, Bethermin20, Barchiesi22}. 

GS\_3073 is compact in all observations, with size estimates in the range $0.08-0.18$~kpc from {\sc{galfit}} fitting \citep{Vanzella10, vdWel12}. Other structural parameters are constrained through {\sc{galfit}} fits to the $H-$band (F160W) data as follows: Sérsic index $n_S=8.00\pm1.86$, axis ratio $q=0.71\pm0.08$, and position angle PA$=75.07^\circ\pm10.79^\circ$, however with a low-quality flag \citep{vdWel12}.

\subsection{{\it JWST}/NIRSpec IFS observations of GS\_3073}\label{s:obs}

GS\_3073 has been observed as part of the NIRSpec IFS GTO program ``Galaxy Assembly with NIRSpec IFS'' (GA-NIFS), under program 1216 (PI: Nora L\"utzgendorf). The target was observed on September $25^{\rm th}$  2022, with a medium cycling pattern of eight dithers and a total integration time of 5~h with the high-resolution grating/filter pair G395H/F290LP \citep[spectral resolution $R\sim1900-3600$;][]{Jakobsen22}. In addition, PRISM/CLEAR observations with a medium cycling pattern of eight dithers and a total integration time of 1.1~h were taken (spectral resolution $R\sim30-330$).

Raw data files were downloaded from the MAST archive and subsequently processed with the JWST Science Calibration pipeline\footnote{\url{https://jwst-pipeline.readthedocs.io/en/stable/jwst/introduction.html}} version 1.8.4 under CRDS context jwst\_1014.pmap. We made several modifications to the default reduction steps to increase data quality, which are described in detail by \cite{Perna23} and which we briefly summarize here. 
Count-rate frames were corrected for $1/f$ noise through a polynomial fit. 
Outliers were flagged on the individual 2-d exposures, using an algorithm similar to {\sc lacosmic} \citep{vDokkum01}. Because the point-spread function (PSF) is under-sampled in the spatial direction across IFU slices, we calculated the derivative of the count-rate maps only along the dispersion direction. The derivative was then normalised by the local flux (or by 3 times the rms noise, whichever was highest), and we rejected the 98\textsuperscript{th} percentile of the resulting distribution (see F.~D'Eugenio et al.\ 2023 for details). 
In place of performing the default pipeline steps {\it flat\_field} and {\it photom} during Stage 2, we performed an external flux calibration after Stage 3 by utilising the commissioning observations of a standard star (program 1128, `Spectrophotometric Sensitivity and Absolute Flux Calibration', PI: Nora L\"utzgendorf; observation 9), reduced under the same CRDS context \citep[see also e.g.][]{Cresci23, Veilleux23}. Uncertainties related to the flux calibration are expected to be of the order of a few percent \citep[see][]{Boeker23, Perna23}. 
The final cube was combined using the `drizzle' method, for which we used an official patch to correct a known bug\footnote{\url{https://github.com/spacetelescope/jwst/pull/7306}}. The main analysis in this paper is based on the combined R2700 cube with a pixel scale of $0.1\arcsec$.
We used spaxels away from the central source and free of emission features to perform a background subtraction.

We use the same steps to reduce the R100 data, though here we explore cube combinations with smaller pixel scales (down to $0.03\arcsec$) for a sharper view of the source morphology particularly in the bluest wavelengths (see Appendix~\ref{a:r100_comp}). A more detailed analysis of the lower-resolution data will be presented in future work.

\section{Analysis}\label{s:analysis}

\subsection{Detection of a BLR and outflow component}

In Figure~\ref{f:3x3spec} we show the integrated G395H spectrum extracted by summing the spectra from the central three by three spaxels ($0.3\arcsec\times0.3\arcsec$, or 1.8~kpc $\times$1.8~kpc), which should encompass the nuclear emission, in the wavelength range $2.86\micron < \lambda < 4.86\micron$. Here, and for the fitting described in Section~\ref{s:fitting}, we calculated an initial noise spectrum by adding the errors in quadrature, i.e. assuming uncorrelated noise between adjacent spaxels based on the error extension in the data cube (‘ERR’). This noise spectrum was then re-scaled with a measurement of the standard deviation in the integrated spectrum in regions free of line emission to take into account correlations due to the non-negligible size of the PSF relative to the spaxel size. This increases the noise by about a factor of three (grey error bars in Figure~\ref{f:3x3spec}).

Several emission lines are present in the G395H spectrum. We note the prominent, very broad component in all permitted lines in the spectrum: H$\beta$, H$\alpha$, He\,{\sc{ii}} and seven He\,{\sc{i}} lines. These broad components of the permitted lines, and the fact that they are not present in the forbidden lines, unambiguously reveal the presence of a Broad Line Region (BLR) around an accreting black hole. The fact that these lines are significantly fainter than their narrow counterparts implies a type 1.8 classification of this AGN \citep[e.g.][]{AGN90, Whittle92}. We report the detection of high-ionisation [Ar\,{\sc{iv}}] and [Ar\,{\sc{iii}}]. With grey dotted vertical lines we indicate the theoretical positions of several coronal lines, of the auroral line [N\,{\sc{ii}}]$\lambda5755$, and of another emission line at $\lambda\sim7167.5$\AA, potentially Si\,{\sc{i}}. 

In addition, the base of the [O\,{\sc{iii}}] doublet reveals the presence of a slightly redshifted broad wing (but much narrower than the BLR permitted components), which we assume to trace a galactic outflow. For a clearer distinction from the BLR and narrow line components, we will call this broad wing the `outflow' component.

\subsection{Fitting of the integrated G395H spectrum}\label{s:fitting}

\begin{figure*}
	\centering
	\includegraphics[width=0.49\textwidth]{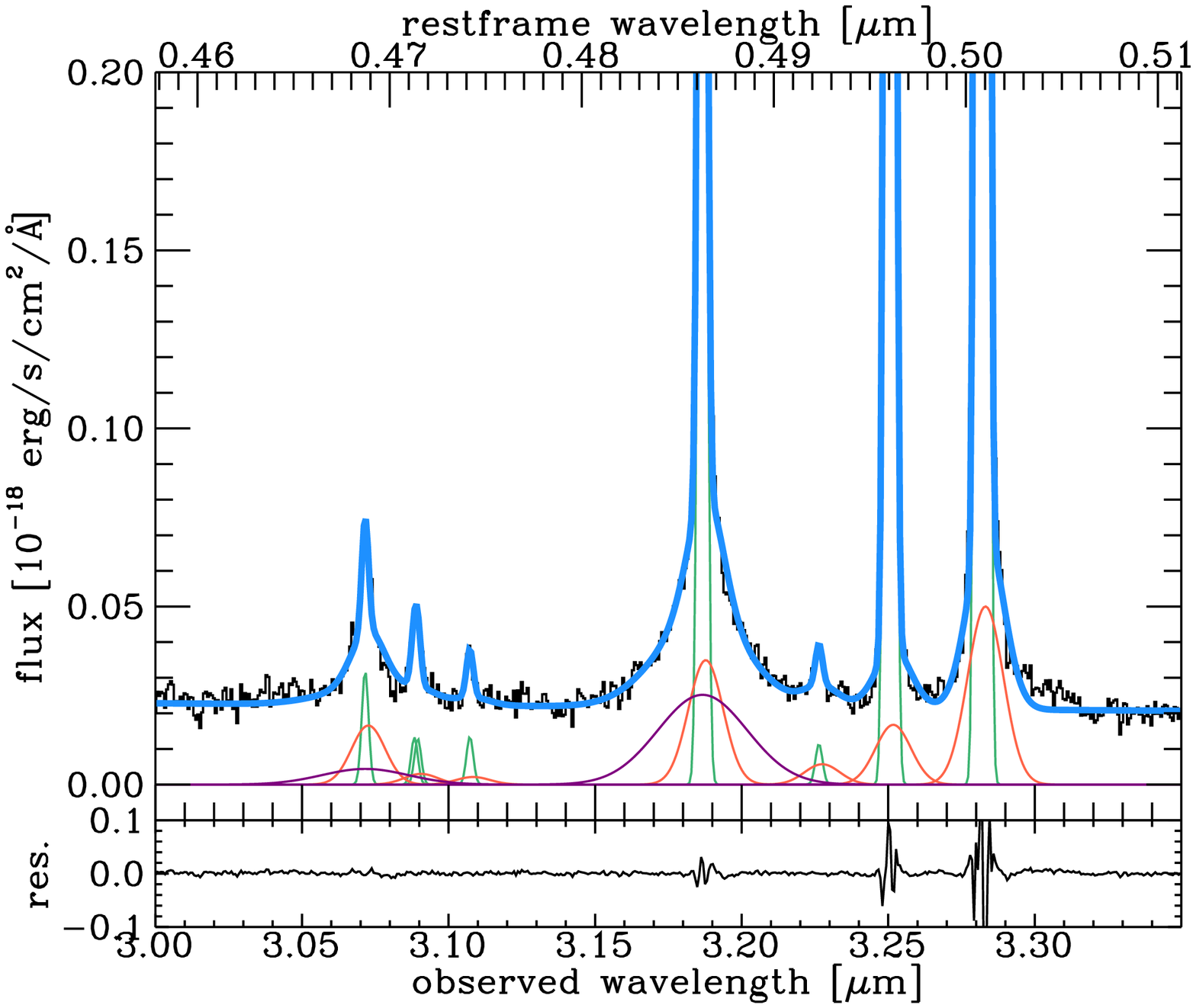}
	\includegraphics[width=0.49\textwidth]{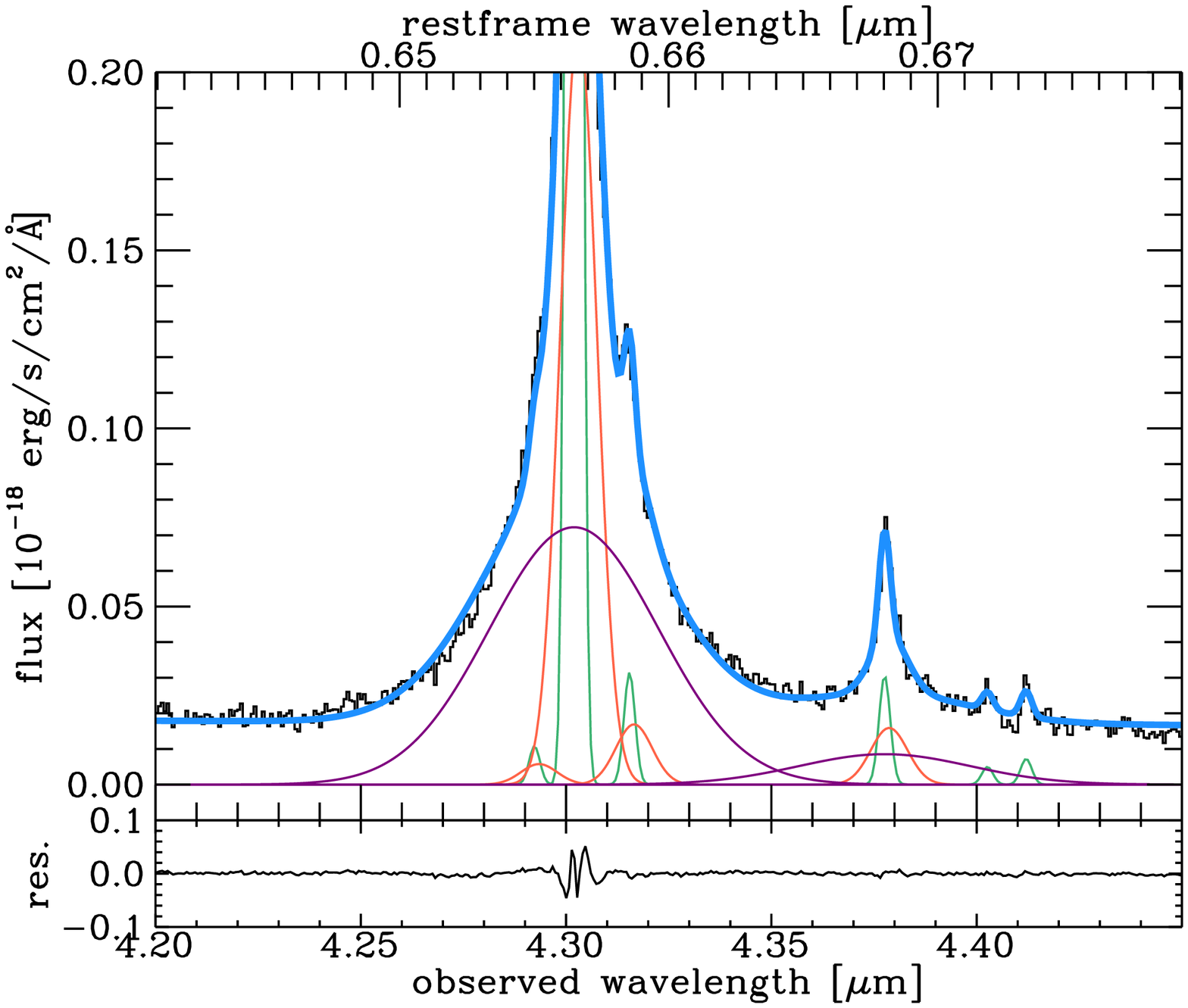}
	\caption{Zoom-in on the integrated spectrum extracted from the central three by three spaxels including our best fit (blue) with individual components for the emission lines (narrow: green, outflow: orange, BLR: purple) for the wavelength range including \heii, \ariv$\lambda4711$, \hei$\lambda4713$, \ariv$\lambda4740$, \hb, \hei$\lambda4922$, \oiii$\lambda4959$, and \oiii$\lambda5007$ (left) and the wavelength range including \nii$\lambda6548$, \ha, \nii$\lambda6584$, \hei$\lambda6678$, \sii$\lambda6716$, and \sii$\lambda6731$ (right). The bottom panels show the residuals, res. = data -- best fit. Note that for some of the weaker lines no BLR or outflow component is preferred by the fit. We further note a faint flux excess red-wards of \oiii$\lambda5007$ that is undetected in individual spaxels, potentially suggesting some higher velocity (nuclear) outflow components not captured by our fiducial fit.}
	\label{f:fit}
\end{figure*}

We fit the full spectrum shown in Figure~\ref{f:3x3spec}: we include a narrow line component tracing the emission from the host galaxy (both from the narrow line region and any star formation) for all lines, a broader line component tracing the outflow emission for all lines, and a BLR component for the hydrogen and helium lines, tracing the high-density gas in the BLR of the AGN. 
We estimate the continuum through a first-order polynomial, independently for the two spectral regions separated by the detector gap around $4.1\micron$. 
For the full fit, we tie the relative velocities for all narrow and BLR lines together, and we fix the velocity of the outflow components relative to the narrow components. 
The velocity widths of all BLR components are tied, assuming that these emission components originate from the same region. We note that this is likely a simplification as different species may be stratified in the BLR \citep{Peterson99}, but this approach allows us to better model the BLR components of the faint helium lines.
For the narrow and outflow lines, we tie the velocity widths separately around \ha\ and around \oiii. 
This allows us to account to first order for the different spectral resolutions around these main line complexes, while still constraining, where possible, the different components of fainter emission lines. 
We have also performed fits tying the line widths in velocity taking into account specifically the linearly wavelength-dependent spectral resolution of the G395H grating. Overall, this gives very similar results in terms of black hole mass estimates and narrow line ratios, however, we clearly achieve a better fit to the narrow [N\,{\sc{ii}}]$\lambda6584$ emission with our fiducial approach. The line widths of the outflow components are most affected by this choice (see Table~\ref{t:fit}). 

The fit is further constrained through atomic physics
\citep[e.g.][]{Osterbrock06}: the intensity ratio of the [O\,{\sc{iii}}] doublet is fixed to [O\,{\sc{iii}}]$\lambda5007$/[O\,{\sc{iii}}]$\lambda4959=2.98$, and the intensity ratio of the [N\,{\sc{ii}}] doublet is fixed to [N\,{\sc{ii}}]$\lambda6584$/[N\,{\sc{ii}}]$\lambda6548=2.94$. 
We verify that the [S\,{\sc{ii}}] and \ariv\ doublet ratios are within the physically allowed ranges of $0.44<$~[S\,{\sc{ii}}]$\lambda6716$/[S\,{\sc{ii}}]$\lambda6731<1.45$ and $0.12<$~[Ar\,{\sc{iv}}]$\lambda4711$/[Ar\,{\sc{iv}}]$\lambda4740<1.37$ \citep[for $T_e=10^4$~K; see][for \ariv]{Proxauf14}. 
We show a zoom-in of our best fit in Figure~\ref{f:fit}. Narrow, outflow, and BLR lines are indicated in green, orange, and purple, and the full fit is shown in blue.

We report emission line properties as constrained through our best fit and used in Sections~\ref{s:measurements} and \ref{s:discussion} in Table~\ref{t:fit}. In addition, we briefly discuss line fluxes of \hei\ in Appendix~\ref{a:hei}.
We use the BLR components to measure the central black hole mass in Section~\ref{s:bhmass}, and the outflow line components to constrain outflow properties in Section~\ref{s:outflow}.
In our discussion of line ratio diagnostics in Sections~\ref{s:bpt} and \ref{s:heii} we quote values based on narrow and outflow line components.

\begin{table}
\caption{Emission line properties of GS\_3073 as constrained through our best fit (Section~\ref{s:fitting}). Uncertainties are calculated from the formal best fit to the R2700 data. Quoted FWHM are corrected for the wavelength-dependent instrumental resolution at the line position. FWHM and velocities are expressed in units of km/s, and line fluxes are expressed in units of $10^{-18}$ erg/s/cm$^2$.}
\begin{tabular}{lc}
\hline
\hline
    Measurement & Value  \\
\hline 
    FWHM$_{\rm H\alpha, narrow}$  & $195.2\pm1.7$ \\
    FWHM$_{\rm H\beta, narrow}$  & $224.7\pm0.6$ \\
    FWHM$_{\rm H\alpha, outflow}$  & $716.6\pm12.0$ \\
    FWHM$_{\rm H\beta, outflow}$$^a$ & $1306.7\pm23.7$ \\
    FWHM$_{\rm H\alpha, BLR}$  & $3370.4\pm41.4$ \\
    $\Delta v_{\rm H\alpha, narrow,outflow}$  & $76.8\pm2.5$ \\
    $\Delta v_{\rm H\beta, narrow,outflow}$  & $100.6\pm4.3$ \\    
\hline    
    $F_{\rm H\alpha,narrow}$  & $38.71\pm0.43$ \\
    $F_{\rm H\alpha,outflow}$  & $23.59\pm0.76$ \\
    $F_{\rm H\alpha,BLR}$  & $37.22\pm0.95$ \\
\hline    
    $F_{\rm H\beta,narrow}$  & $16.04\pm0.26$ \\
    $F_{\rm H\beta,outflow}$  & $5.21\pm0.35$ \\
    $F_{\rm H\beta,BLR}$  & $9.61\pm0.48$ \\
\hline
    $F_{\rm He\,II,narrow}$  & $0.93\pm0.09$ \\
    $F_{\rm He\,II,outflow}$  & $2.38\pm0.32$ \\
    $F_{\rm He\,II,BLR}$  & $1.61\pm0.47$ \\ 
    EW(HeII$_{\rm narrow+outflow}$) [\AA] & $20.42\pm3.09$ \\
\hline    
    $F_{\rm [O\,III]\lambda5007,narrow}$  & $135.94\pm0.41$ \\ 
    $F_{\rm [O\,III]\lambda5007,outflow}$  & $7.69\pm0.14$ \\   
    $F_{\rm [N\,II]\lambda6584,narrow}$  & $1.07\pm0.11$ \\
    $F_{\rm [N\,II]\lambda6584,outflow}$  & $1.88\pm0.25$ \\   
    $F_{\rm [S\,II]\lambda6717+\lambda6731,narrow}$  & $0.42\pm0.08$ \\
\hline   
    $T_{e,\rm [O\,III]}\ [K]^{b}$ & 14163$^{+1339}_{-1439}$ \\
\hline
    \multicolumn{2}{p{1.0\columnwidth}}{$^a$FWHM$_{\rm H\beta, outflow}$ = FWHM$_{\rm [O\,III], outflow}$
    
    $^{b}$$T_{e,\rm [O\,III]}$ is constrained from the narrow line ratio of [O\,{\sc{iii}}]$\lambda4363$/([O\,{\sc{iii}}]$\lambda4959$+[O\,{\sc{iii}}]$\lambda5007$). We emphasize that the [O\,{\sc{iii}}]$\lambda4363$ emission is only partially covered by our spectrum, but the fit is constrained though the line positions and widths of other high-$S/N$ lines.}
\end{tabular} 
\label{t:fit}
\end{table}

\subsection{Kinematic maps}\label{kinmaps}

\begin{figure*}
	\centering
	\includegraphics[width=0.25\textwidth]{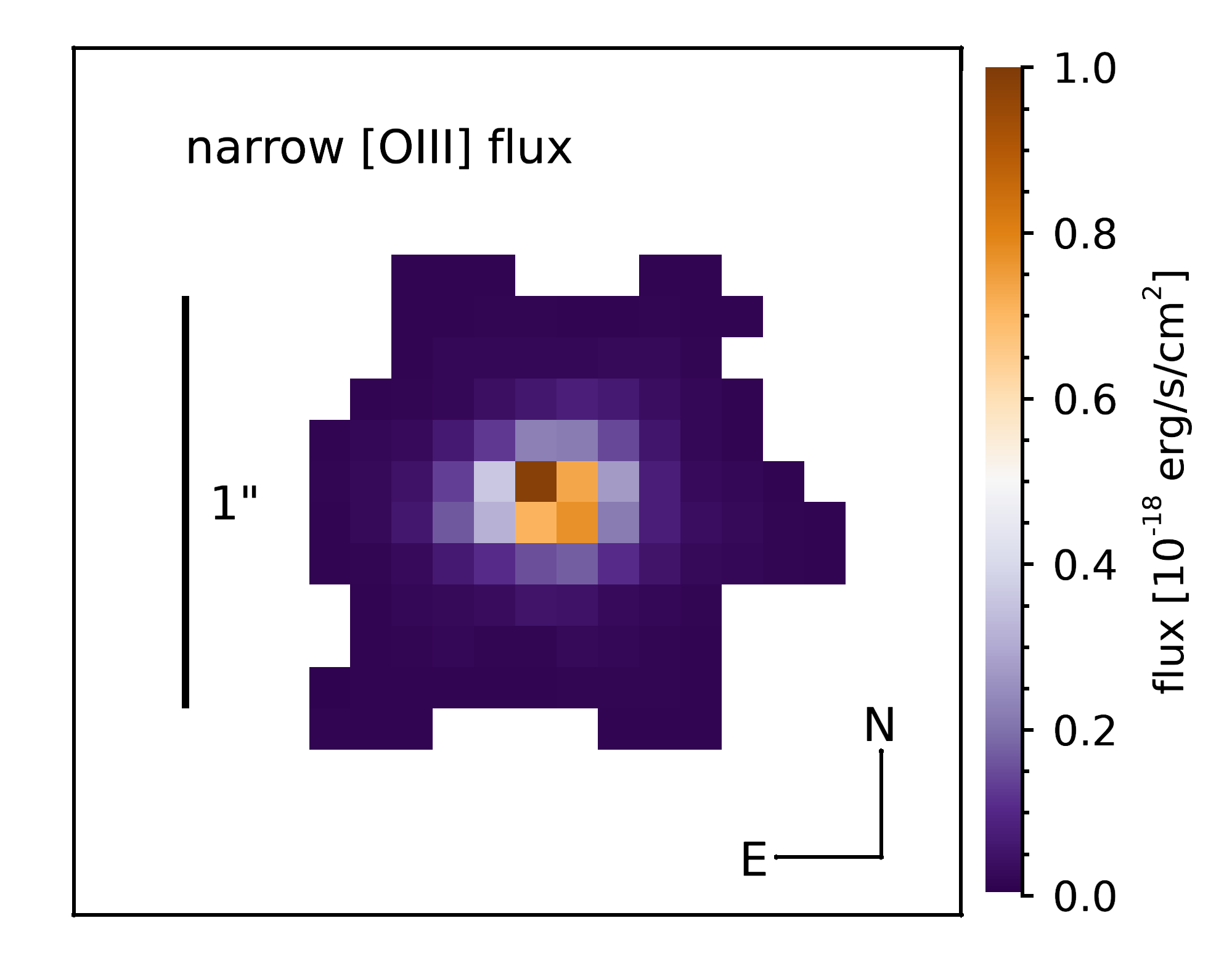}
	\includegraphics[width=0.25\textwidth]{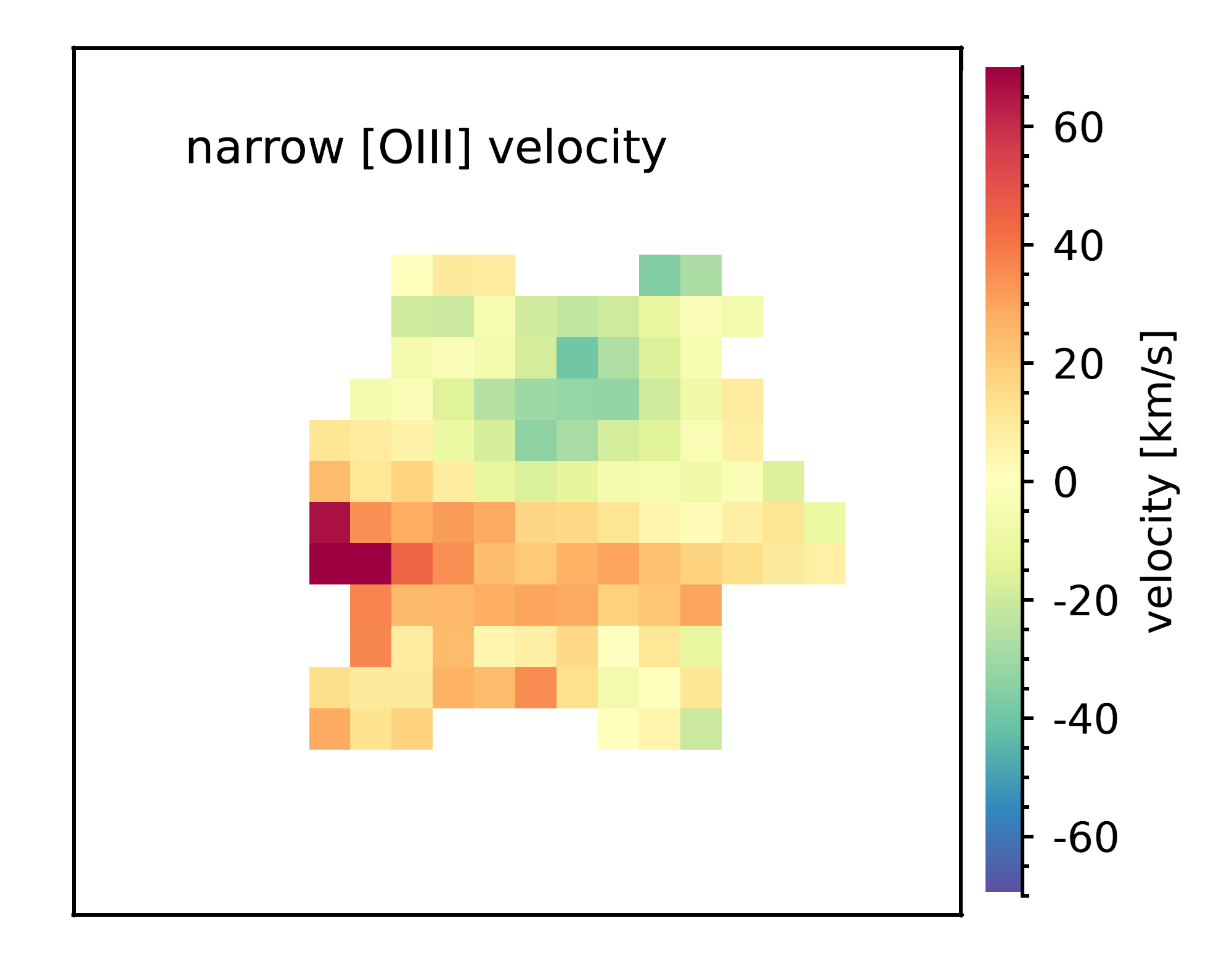}
	\includegraphics[width=0.25\textwidth]{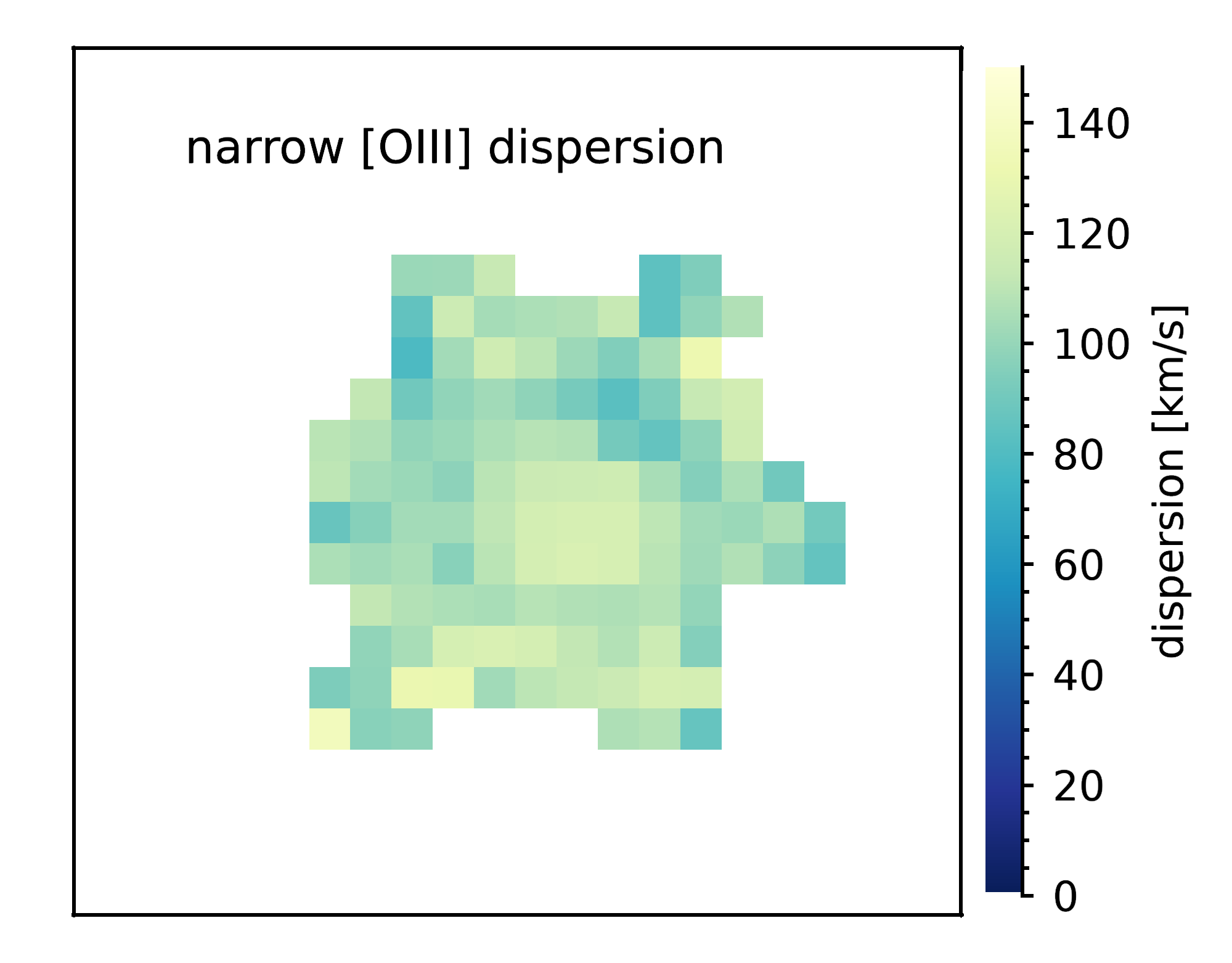}	
	\includegraphics[width=0.25\textwidth]{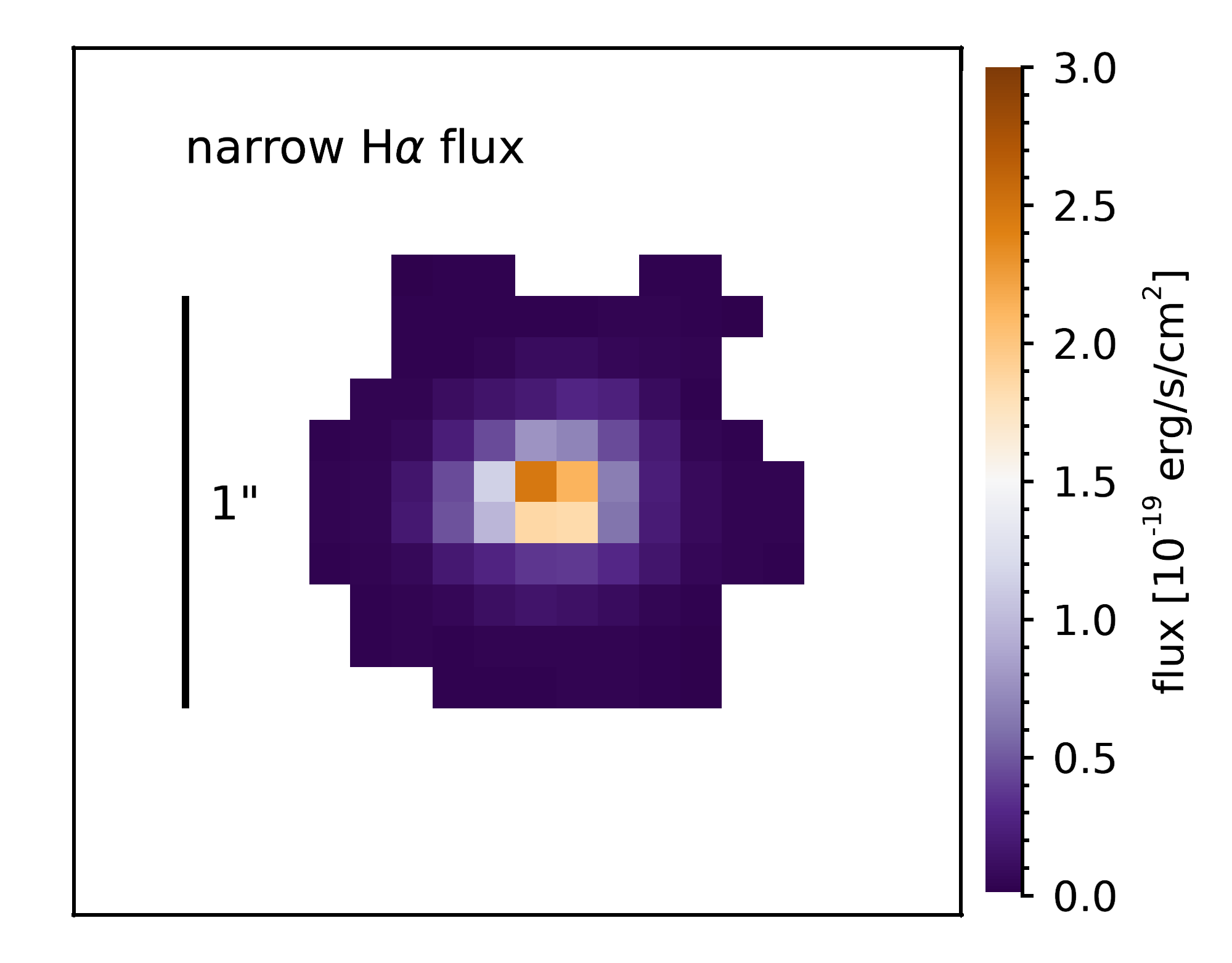}
	\includegraphics[width=0.25\textwidth]{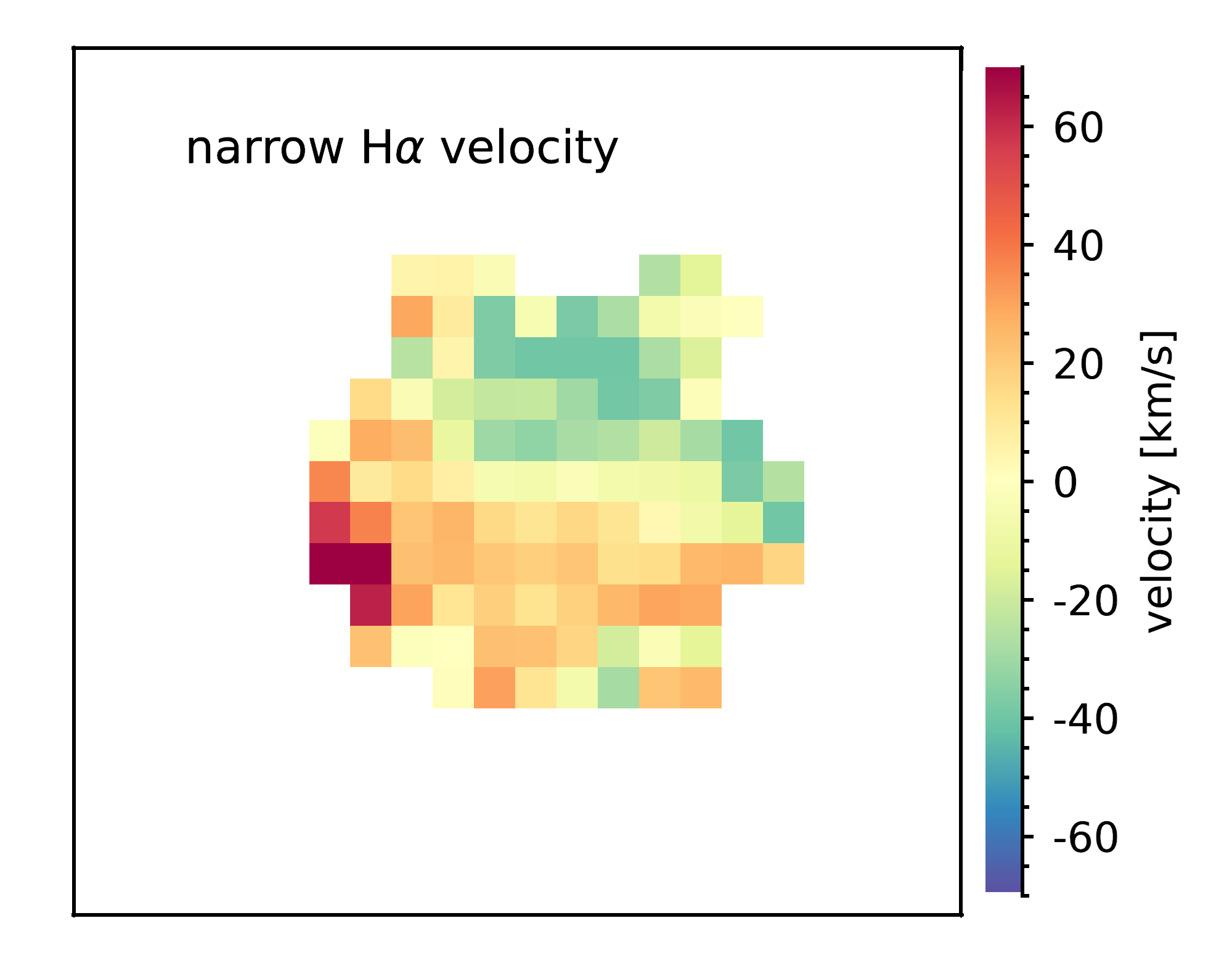}
	\includegraphics[width=0.25\textwidth]{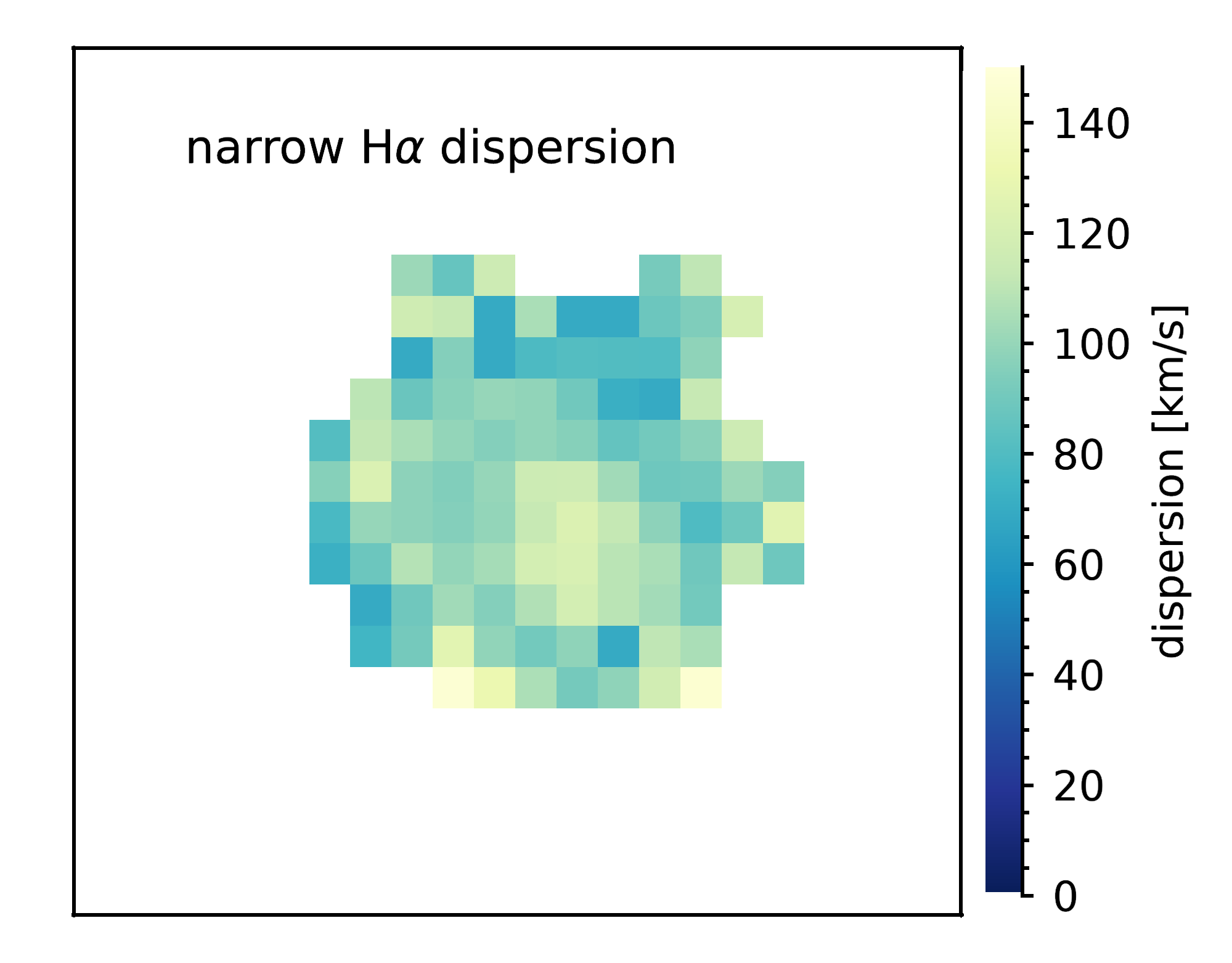}
	\includegraphics[width=0.25\textwidth]{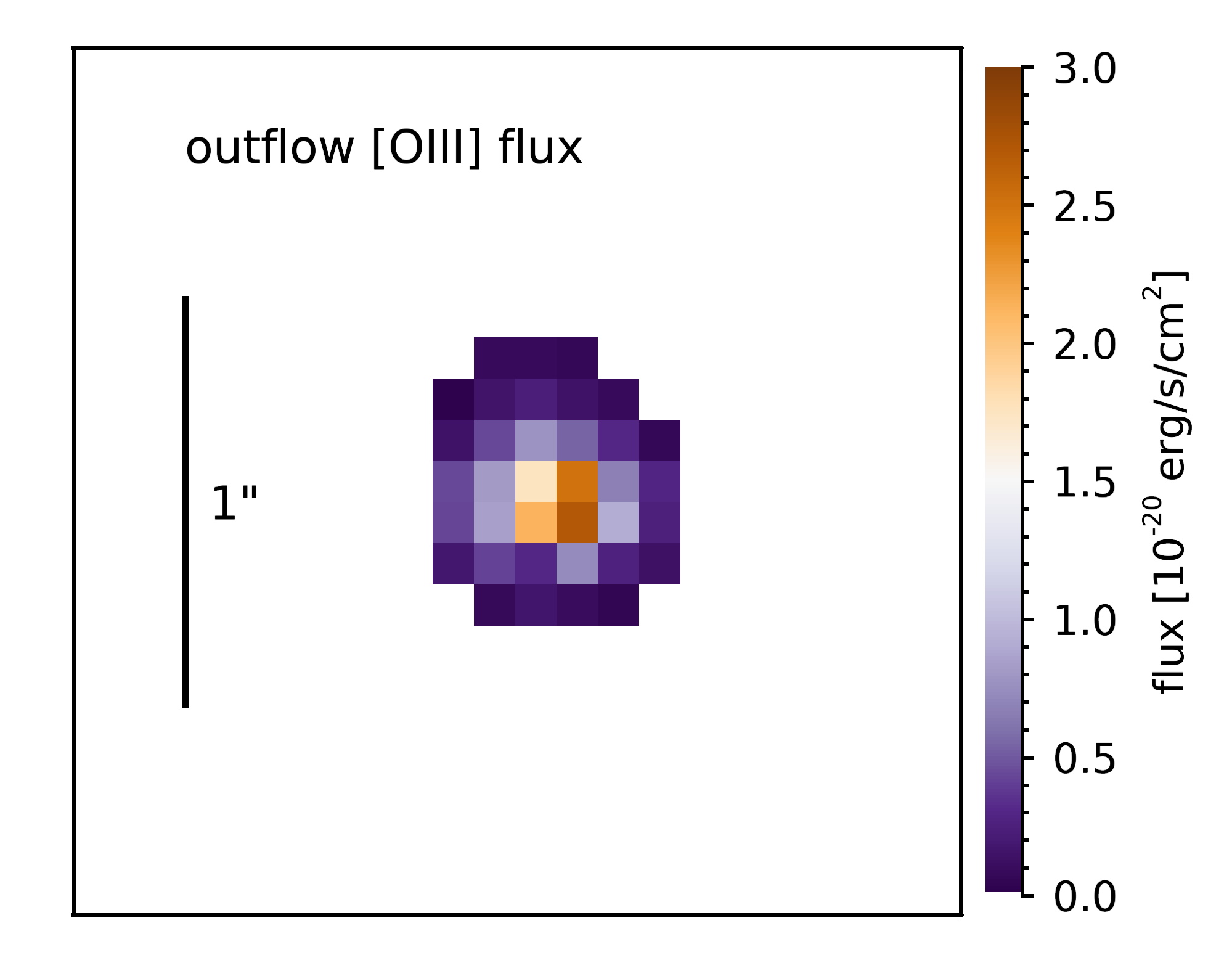}
	\includegraphics[width=0.25\textwidth]{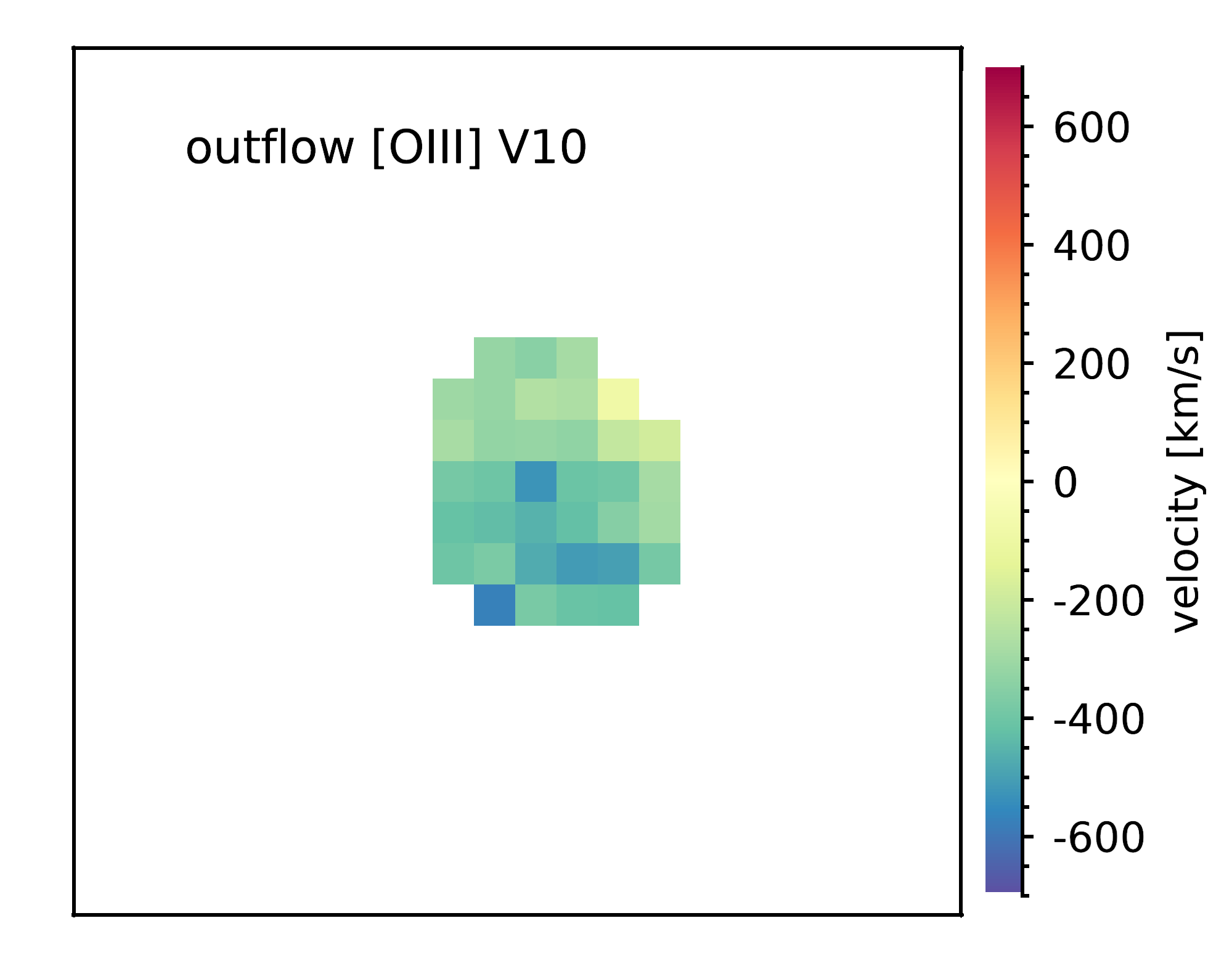}
	\includegraphics[width=0.25\textwidth]{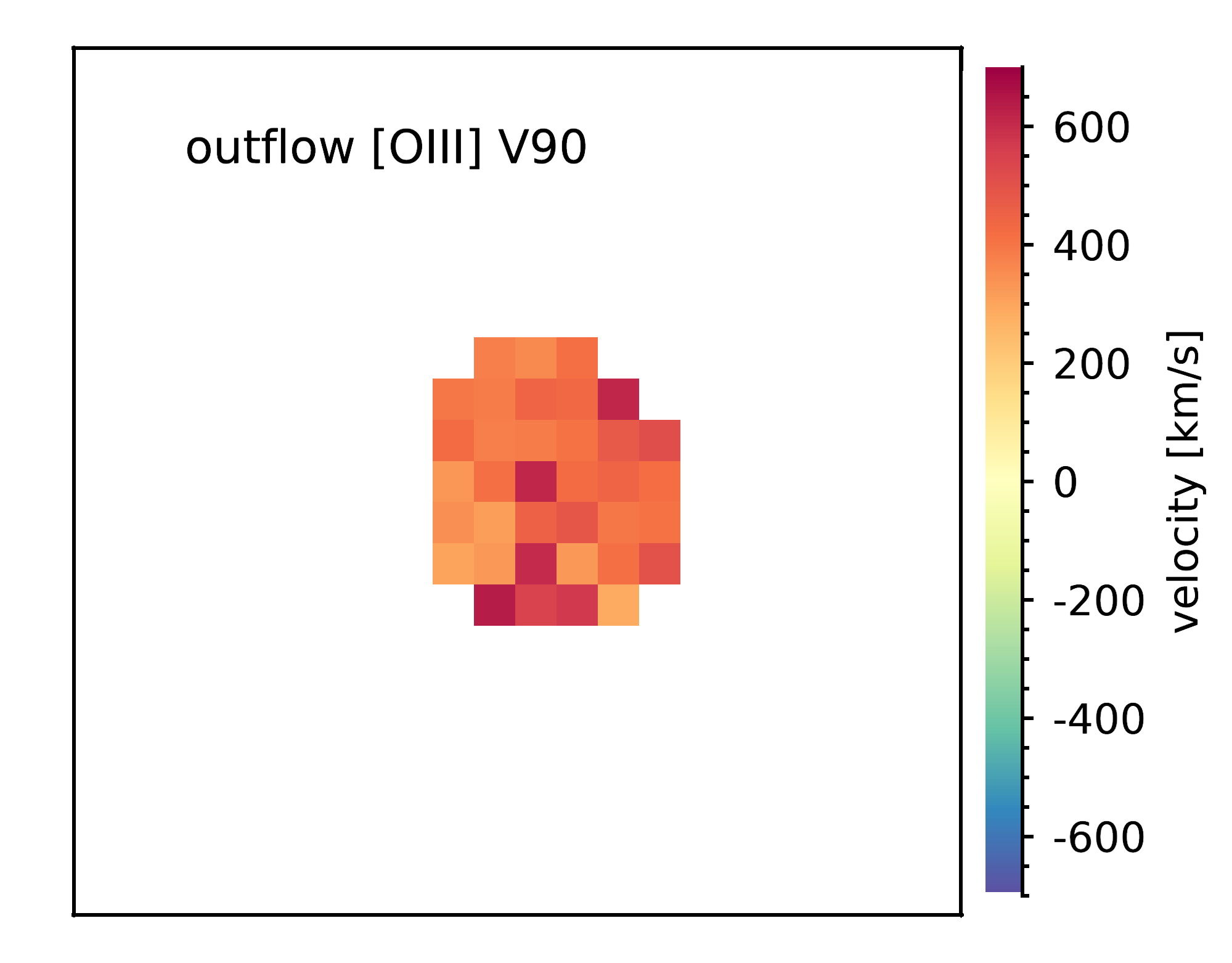}		
    \caption{Top and middle row: projected maps of flux (left), velocity (middle), and velocity dispersion corrected for instrumental resolution (right) as measured from the narrow [O{\sc{iii}}] component (top) and the narrow H$\alpha$ component (middle) tracing the host galaxy kinematics. Bottom: projected maps of flux (left), V10, the velocity at the 10$^{th}$ percentile of the emission-line profile (middle), and V90, the velocity at the 90$^{th}$ percentile (right), from the outflow [O{\sc{iii}}] component. North is up and East is to the left. The spaxel size is $0.1\arcsec$. A bar in the first column indicates $1\arcsec$, corresponding to roughly 6~kpc at the source redshift. The highest projected velocities are found in the East-South-East region in both the narrow \ha\ and \oiii\ velocity maps, and are possibly related to an ongoing interaction with a faint companion (see Appendix~\ref{a:r100_comp}). Apart from this region, a velocity gradient of $\Delta v\sim70$~km/s along the North-North-West to South-South-East direction is visible in both the narrow [O{\sc{iii}}] and \ha, possibly indicative of rotation.
    There is a small offset between the integrated [O{\sc{iii}}] and H$\alpha$ line centroids of about 8~km/s. We find elevated velocity dispersions in the galaxy centre, as expected for observations of a rotating disc affected by beam-smearing. 
    The outflow is visible in the nuclear region, with positive and negative velocities respectively the systemic velocity of |600-700|~km/s.
    }
	\label{f:kinmaps}
\end{figure*}

We derive maps of projected flux, velocity, and velocity dispersion for the narrow \oiii\ and \ha\ lines, and for the outflow component as traced by \oiii\ from multi-component Gaussian fits to the continuum-subtracted emission. Specifically, to derive the 2d maps we fit the emission in each spaxel separately around \oiii\ and around \ha. We use two components for \oiii, one for the narrow line emission and one for the broader component which we interpret as an outflow. The amplitude, width and line centroid of both components are free to vary in these fits. For the fit of the \ha\ complex we also use two components, one for the narrow line emission and one for the BLR emission. While we fix the line centroid of the BLR emission to the narow line emission, the amplitude and width of both components are free to vary. This simplified approach (i.e.\ without explicitly accounting for an outflow component or \nii\ emission around \ha) allows us to fit the lower$-S/N$ spectra from individual spaxels and in particular at larger distances from the source centre. However we emphasize that the narrow \ha\ centroid and width are still well constrained due to the outflow and BLR emission being much fainter. We visually inspect the fits to each spaxel and create masks accordingly. 
The resulting maps are shown in Figure~\ref{f:kinmaps}.

A velocity gradient of $\Delta v\sim70$~km/s roughly along the North-South direction is visible in both the narrow [O\,{\sc{iii}}] and H$\alpha$ components, which may be interpreted as rotation.
There is a small offset between the integrated [O\,{\sc{iii}}] and H$\alpha$ line centroids of about 8~km/s, much lower than the spectral resolution of our data. Note that we find a kinematic position angle of about PA$_{\rm kin}\sim-15^\circ$, perpendicular to the morphological PA$=75^\circ$ (see Section~\ref{s:ancillary}).
However, we also note that this compact source is barely resolved in the combined cube with $0.1\arcsec$ pixel scale, as indicated by the point-spread function (PSF) intensity profile imprinted on our line maps (see Figure~\ref{f:kinmaps}). 

Going to a smaller plate scale and bluer wavelengths in the R100 data, we do see some evidence for structure in the main galaxy, and possibly even for two faint companions (see Appendix~\ref{a:r100_comp}). 
There is a region of higher positive velocities visible in both the \ha\ and \oiii\ maps in the East-South-East direction. It is conceivable that this kinematic feature is related to one of these faint tentative companion galaxies. 

The velocity dispersion maps are relatively featureless, but we see elevated dispersions in the central region of the galaxy in both the \ha\ and \oiii\ maps. This is expected for the observed kinematics of a rotating disc that are affected by beam-smearing particularly in the centre. We do not see an increase in dispersion toward the region of higher velocities in the East-South-East.

The outflow is visible in the nuclear region, with positive and negative velocities measured from V10 and V90 of up to |600-700|~km/s (with respect to the systemic velocity of the galaxy; see bottom middle and right panels in Figure~\ref{f:kinmaps}).

\section{Results}\label{s:measurements}

Physical properties derived from our best fit to the fiducial spectrum integrated over the central three by three spaxels are reported in Table~\ref{t:properties}. We describe our measurements here.

\begin{table}
\caption{Measurements of central black hole mass, host galaxy dynamical mass, and outflow properties (see Section~\ref{s:measurements}).}
\begin{tabular}{lc}
\hline 
\hline
    Measurement & Value  \\
\hline 
    $\log(M_{\rm BH}/M_\odot)_{\rm H\alpha}$ & $8.2\pm0.4$ \\
    $\log(L_{\rm bol}/(\rm erg/s))$ & $\sim45.2-46.2$ \\
    $\log(L_{\rm Edd}/(\rm erg/s))$ & $46.3\pm0.4$ \\
    $\lambda_{\rm Edd}$ & $\sim0.1-1.6$ \\
\hline
    $\log(M_{\rm dyn}/M_\odot)$ & $9.4^{+0.7}_{-0.2}$ \\
\hline
    $v_{\rm out,H\alpha}=\langle v_{\rm outflow}\rangle+2\sigma_{\rm outflow}$ [km/s] & $685\pm11$\\
    $\dot{M}_{\rm out,ion}$ [$M_\odot$/yr] & $98\pm2$ \\
    $\eta_{\rm ion}$ & $6.1^{+5.0}_{-2.9}$ \\
\hline
\end{tabular} 
\footnotesize{}
\label{t:properties}
\end{table}

\subsection{Black hole properties}\label{s:bhmass}
 
We find a H$\alpha/$H$\beta$ flux ratio of $3.87\pm0.22$ for the BLR components. For QSOs, intrinsic H$\alpha/$H$\beta$ ratios of $3-10$ are routinely observed \citep[e.g.][]{Osterbrock77, Osterbrock81}, and therefore no correction for extinction is performed. 
If the BLR \ha\ luminosity were underestimated due to the presence of dust, the black hole mass calculated below would correspond to a lower limit.

Assuming that the gas in the BLR is virialized, we calculate the central black hole mass from the spectral properties of the H$\alpha$ BLR region following the calibration by \cite{Reines13}:\footnote{
We remind the reader that this calibration has not yet been thoroughly tested at high redshift, because rest-frame optical lines were not accessible at z>4 before the launch of {\it JWST}.}

\begin{multline}\label{eq:mbh}
\log\left(\frac{M_{\rm BH}}{M_\odot}\right)=\log(\epsilon) + 6.57 \\ + 0.47 \log\left(\frac{L_{\rm H\alpha}}{10^{42}{\rm erg/s}}\right) + 2.06 \log\left(\frac{\rm FWHM_{\rm H\alpha}}{10^{3}{\rm km/s}}\right)
\end{multline}
(their Equation 5; see also \citealp{Greene05}), where $\epsilon$ is a scaling factor depending on the structure, kinematics, and orientation of the BLR, of the order of $\sim1$, and reported in the range $0.75-1.4$ \citep{Onken04, Reines13}. 
Following \cite{Reines15}, we therefore assume $\epsilon=1.075\pm0.325$.
The luminosity $L_{\rm H\alpha}$ and FWHM$_{\rm H\alpha}$ and their uncertainties are measured from our best-fit BLR component.
In addition, we account for the statistical uncertainty on $\epsilon$ as based on local $M_{\rm BH}-\sigma_\star$ relations by adding 0.4~dex in quadrature to our error budget \citep[e.g.][]{Ho14}, which dominates the uncertainty.  We measure a black hole mass of $\log(M_{{\rm BH}}/M_{\odot})=8.2\pm0.4$ (Table~\ref{t:properties}).

We verify that using a larger aperture (while constraining the BLR width to the best fit from the fiducial higher signal-to-noise spectrum) does not change our results on the black hole mass significantly: by integrating the spectrum over the central $1.15\arcsec$ we find $\log(M_{{\rm BH}}/M_{\odot})=8.3\pm0.4$. This indicates that the BLR emission is nearly fully encompassed by the central three by three spaxels.

Using alternatively the calibrations by \cite{Greene05} utilising the H$\alpha$ BLR emission, the H$\beta$ BLR emission, and the $L_{5100}$ luminosity, and again accounting for the scatter of the local virial relations, we find $\log(M_{{\rm BH}}/M_{\odot})=8.0\pm0.4$, $\log(M_{{\rm BH}}/M_{\odot})=7.9\pm0.4$, and $\log(M_{{\rm BH}}/M_{\odot})=8.2\pm0.4$, respectively, consistent with the estimate using the calibration by \cite{Reines13}.

We use different calibrations to estimate the bolometric luminosity $L_{\rm bol}$ of GS\_3073.\footnote{
Again, we caution that these calibrations are all based on data from low-redshift AGN in the Sloan Digital Sky Survey (SDSS).
}
First, assuming that the narrow line emission is dominated by the AGN, we calculate the $L_{\rm bol}$ from the narrow line luminosities of \hb\ and \oiii\ following Equation~1 by \cite{Netzer09}.
This gives $\log(L_{\rm bol}/({\rm erg/s}))=46.2$, and likely represents an upper limit. If instead we use Equation~25 by \cite{DallaBonta20} to estimate $L_{\rm bol}$ from the \hb\ BLR luminosity, we get $\log(L_{\rm bol}/({\rm erg/s}))=45.3$. We find a similar value of $\log(L_{\rm bol}/({\rm erg/s}))=45.2\pm0.4$ when calculating $L_{\rm bol}$ using Equation~6 by \cite{Stern12} utilising the \ha\ BLR luminosity.

Using our black hole mass estimate from the \ha\ BLR (Eq.\,\ref{eq:mbh}), the Eddington luminosity is $\log(L_{\rm Edd}/({\rm erg/s}))=4\pi G M_{\rm BH} m_{\rm p} c/\sigma_{\rm T}=46.3\pm0.4$, where $G$ is the gravitational constant, $m_{\rm p}$ the proton mass, $c$ the speed of light, and $c_{\rm T}$ the Thomson scattering cross-section. 
Depending on which bolometric luminosity and which black hole mass estimate we adopt, we find Eddington ratios in the range $\lambda_{\rm Edd}=L_{\rm bol}/L_{\rm Edd}=0.1-1.6$  (Table~\ref{t:properties}).

\subsection{Host galaxy dynamical mass}\label{s:mdyn}

\begin{figure*}
	\centering
	\includegraphics[width=0.49\textwidth]{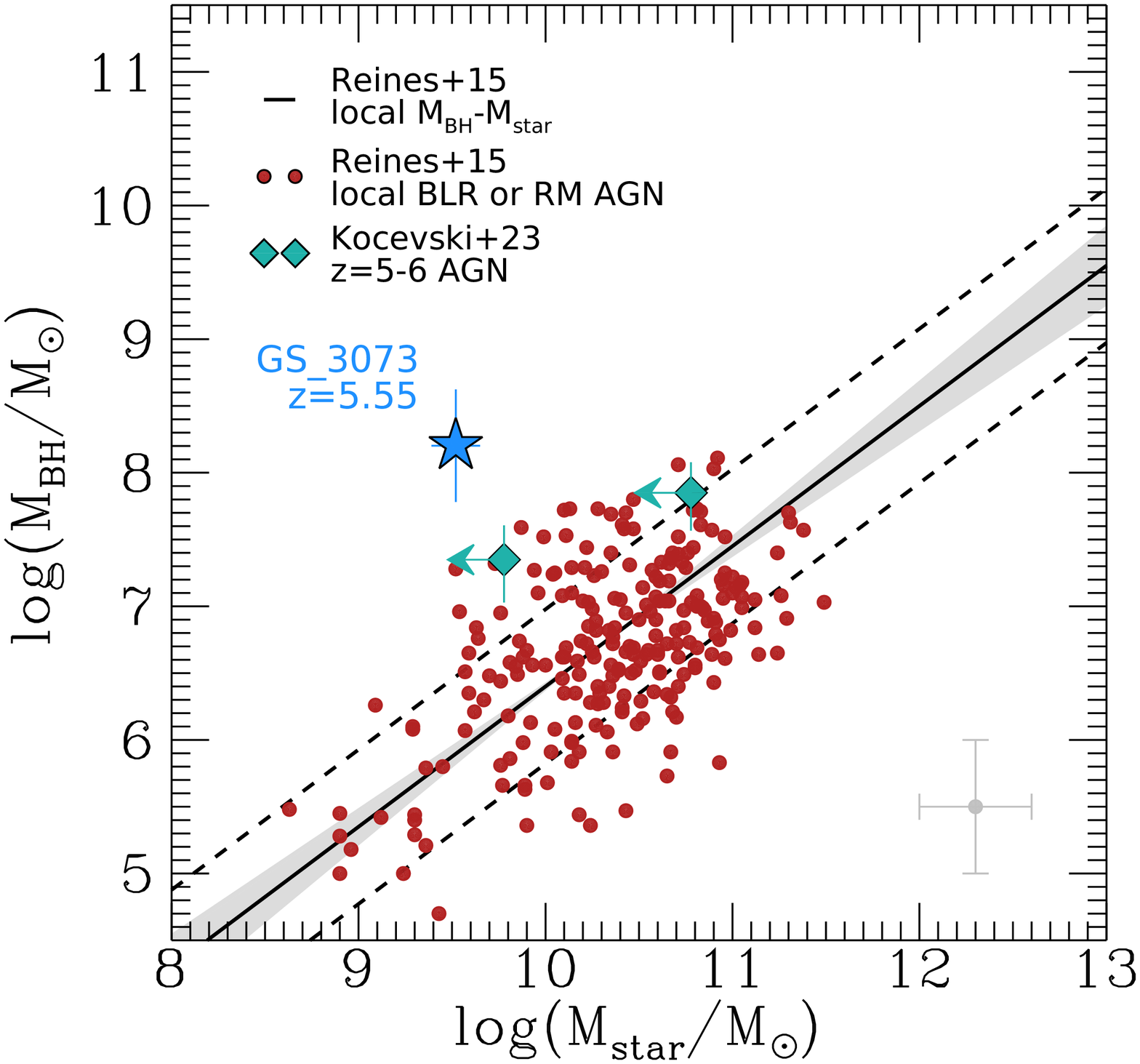}
    \includegraphics[width=0.49\textwidth]{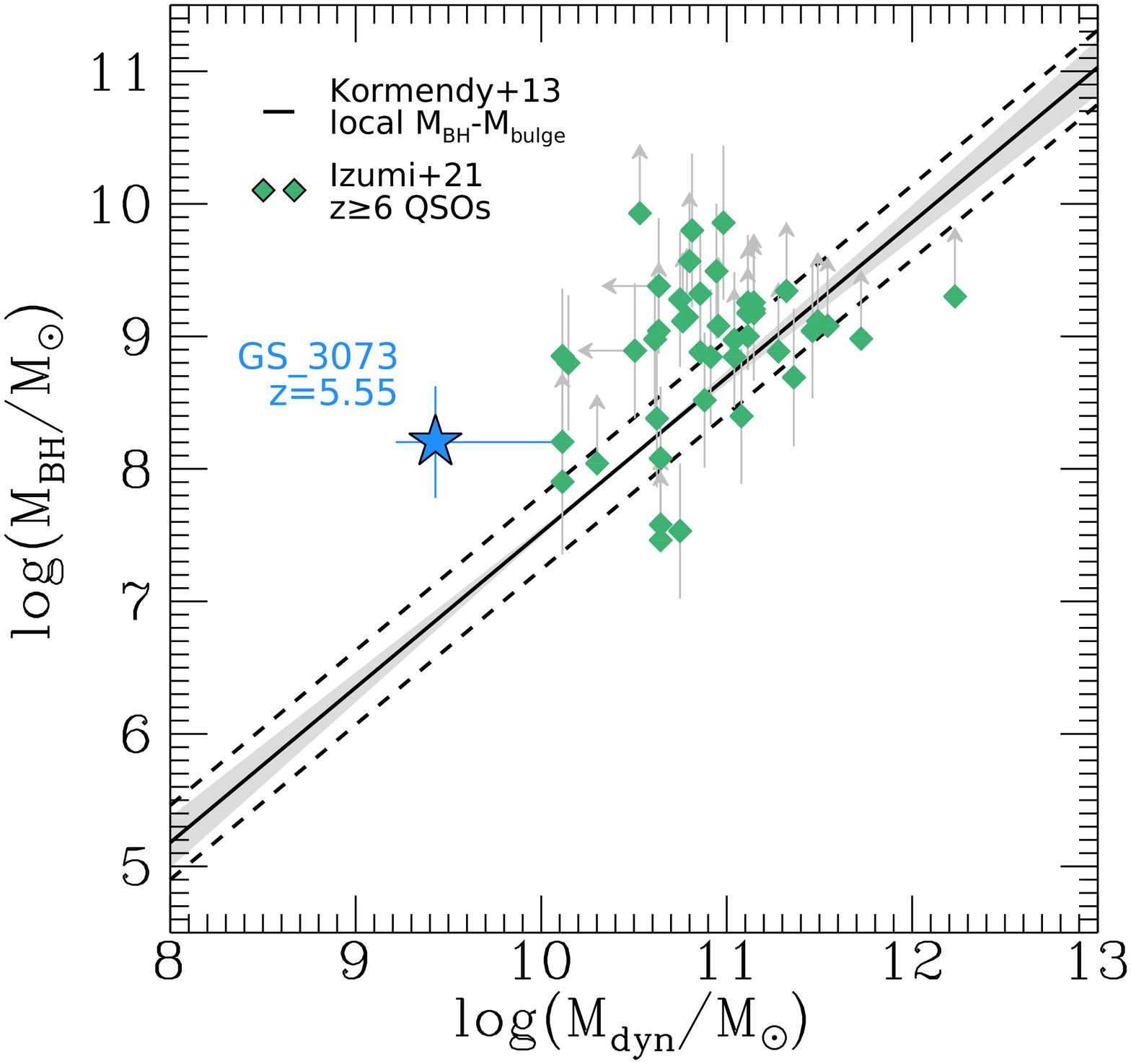}
	\caption{
    Central black hole mass $M_{{\rm BH}}$ as a function of host galaxy stellar mass $M_{\rm star}$ (left) and host galaxy dynamical mass $M_{\rm dyn}$ (right) for GS\_3073 (filled blue star) and literature compilations. In the left panel, we compare our galaxy to local AGN by \citet{Reines15} for which the black hole mass has been determined from a broad line region or from reverberation mapping (red circles, with representative error bar in grey). The black line with shaded region shows the best fit with uncertainties to this data by \citet{Reines15}, with the dashed lines indicating intrinsic scatter plus measurement uncertainties. As teal diamonds we show two data points by \cite{Kocevski23}, for which we calculate the black hole mass based on Eq.~\ref{eq:mbh} for consistency with our measurement and the $z=0$ data.
    In the right panel, we compare our galaxy to $z\geq6$ QSOs by \citet{Izumi21} (green diamonds). For our dynamical mass estimate, we adopt Sérsic index $n_S=4$ and effective radius $R_e=0.18$~kpc, and indicate uncertainties by varying $R_e$ between 0.11 and 0.9~kpc. The black line shows the local $M_{\rm BH}-M_{\rm bulge}$ relation with uncertainties and intrinsic scatter by \citet{Kormendy13}. 
    The black hole of GS\_3073 is at the lower end of the $M_{\rm BH}-M_{\rm dyn}$ distribution when compared to the measurements by \citet{Izumi21}, but still appears over-massive for its host galaxy dynamical mass compared to the local relation by \cite{Kormendy13}. Considering the $M_{\rm BH}-M_{\rm star}$ relation, the black hole of GS\_3073 is much more massive compared to local broad line AGN with a similar host galaxy stellar mass.}
	\label{f:mbh}
\end{figure*}

GS\_3073 is very compact and barely resolved in our IFU observations \citep[see also discussion by][]{Vanzella10}. We estimate its dynamical mass by means of the integrated narrow component line width (corrected for instrumental broadening) as follows:

\begin{equation}\label{eq:mdyn}
    M_{\rm dyn} = K(n)K(q)\frac{\sigma^2 R_e}{G},
\end{equation}

where $K(n)=8.87-0.831n+0.0241n^2$ with Sérsic index $n$ following \cite{Cappellari06}, $K(q)=[0.87+0.38e^{-3.71(1-q)}]^2$ with axis ratio $q$ following \cite{vdWel22}, and $R_e$ is the effective radius.
In this calibration, $\sigma$ is the integrated {\it stellar} velocity dispersion. \cite{Bezanson18b} show that galaxies with low integrated ionised gas velocity dispersion tend to underestimate the integrated stellar velocity dispersion (their Figure~4b). We take this into account with a correction of $\Delta\log(\sigma/({\rm km/s}))=+0.1$ to our measured narrow line dispersion ($\sigma_{\rm H\alpha,narrow}=83$~km/s).

Adopting structural parameters $n=8$, $q=0.71$, and $R_e=0.18$~kpc from Sérsic fits to the $H-$band photometry by \cite{vdWel12}, we find $\log(M_{\rm dyn}/M_\odot)\sim9.2$. Since a Sérsic index of $n=8$ is at the edge of the explored parameter space in the fits by \cite{vdWel12}, likely biased by the presence of the AGN at the centre of the galaxy, we adopt $n=4$ to derive a fiducial dynamical mass, leading to $\log(M_{\rm dyn}/M_\odot)\sim9.4$.
We find similar values when we adopt the calibration between inclination-corrected integrated line widths and disc velocities by \cite{Wisnioski18} together with their equation~3, namely $\log(M_{\rm dyn}/M_\odot)\sim8.9$, or when we attempt to replace the integrated ionised gas velocity dispersion with $v_{\rm rms}^2=v_{\rm obs}^2+\sigma_{\rm obs}^2$, namely $\log(M_{\rm dyn}/M_\odot)\sim9.3$, where $v_{\rm obs}$ is half the inclination-corrected, maximum observed velocity gradient, and $\sigma_{\rm obs}$ is the average observed velocity dispersion of individual spaxels in the outer region of the galaxy (both uncorrected for beam-smearing; see Figure~\ref{f:kinmaps}).
A large uncertainty in these calculations stems from the structural parameters of GS\_3073. Overall, if we vary the size between 0.11 and 0.9~kpc, we find dynamical mass estimates in the range $9.2<\log(M_{\rm dyn}/M_{\odot})<10.1$ (see Table~\ref{t:properties}), and if we vary additionally the Sérsic index between 0.5 and 8 \citep[see values reported by][]{Vanzella10}, we find $9.0<\log(M_{\rm dyn}/M_{\odot})<10.3$.

We note that previous literature estimates on the total stellar mass from SED fitting for GS\_3073 are larger than our dynamical mass estimate, with the most recent estimate being $\log(M_\star/M_\odot)\sim10.6$ \citep{Barchiesi22}. 
The literature estimates are based on broad-band photometry that had unknown emission line contributions.  We find strong emission lines that would significantly contaminate the 3.6 and 4.5~$\mu$m \textit{Spitzer} IRAC (Infra-Red Array Camera) fluxes, the rest-frame optical constraints closest to the Balmer-break region (and therefore most important for stellar mass estimates).  If not accounted for properly, they could lead to over-estimates of the stellar mass.  We used the R100 spectrum to provide an updated stellar mass estimate based on the continuum emission.  We see no strong Balmer-break in the spectrum, and cannot rule out significant contribution, or even dominance of the continuum light from the accretion disc surrounding the black hole.  Under the assumption that the continuum light is dominated by the host galaxy, we fit the continuum in the region $2.73-5.30\mu$m, using \textsc{beagle} \citep{Chevallard16} while fully masking the emission lines (which have multiple physical contributions).  Using a constant star-formation history (SFH) we find a stellar mass estimate of $\log(M_\star/M_\odot)\sim9.52\pm0.13$, where the uncertainties denote the $1\sigma$ credible interval. We find a similar stellar mass estimate when fitting to the full spectrum (rest-frame UV to optical) with a delayed SFH with last 10~Myr of constant star formation allowed to vary independently.
This estimate is consistent within the uncertainties with our dynamical mass estimate, and we use it as a fiducial value for the stellar mass of GS\_3073.

In Figure~\ref{f:mbh} we compare our black hole mass measurement as a function of host galaxy stellar mass and host galaxy dynamical mass to various literature compilations and relations, locally and at high redshift. \cite{Reines15} provide a compilation of local AGN for which the black hole mass has been measured from \ha\ BLR components with the same calibration that we use for our $M_{\rm BH}$ measurement (in addition, some local data points come from reverberation mapping). The black hole of GS\_3073 is more massive by more than two orders of magnitude compared to the best-fit relation to the local sample, and more massive than all local Broad Line AGN by \cite{Reines15}.
We also show two data points from \cite{Kocevski23} at $z\sim5.2$ and $z\sim5.6$ \citep[see also][]{Onoue23}. 
For consistency with our measurement and the $z=0$ data, we re-calculate the black hole masses based on Eq.~\ref{eq:mbh}. This re-calibration results in an increase of the black hole mass by $\sim0.2$~dex, relative to what is quoted by the authors (following the discussion in that paper, we assume $A_V=4$ for the higher mass source).
These two sources appear more consistent with the local BLR population, but since their stellar mass estimates are upper limits, also these black holes could be overly massive.

At high redshift, existing black hole mass measurements are primarily obtained from luminous quasars (QSOs; $\log(L_{\rm bol}/(\rm erg/s))\sim46.5-48$) for which stellar mass estimates are not available. In the right panel of Figure~\ref{f:mbh} we compare the black hole mass of GS\_3073 as a function of host galaxy dynamical mass to $z\geq6$ QSOs compiled by \cite{Izumi21} (see also their figure 13), including data by \cite{Willott10, DeRosa14, Kashikawa15, Venemans15, Banados16, Jiang16, Shao17, Mazzucchelli17, Decarli18}. Our galaxy sits at the lower end of the QSO $M_{{\rm BH}}$ distribution \citep[see also the compilations by][]{Willott17, Pensabene20}. Similar to several of the measurements compiled by \cite{Izumi21}, GS\_3073 appears to sit above the local $M_{{\rm BH}}-M_{\rm bulge}$ relation constrained by \cite{Kormendy13}. 
However, we note that the comparison of the high$-z$ dynamical mass measurements to the local bulge mass measurements, constrained from elliptical and S/S0 galaxies, is not straight forward. As discussed by \cite{Reines15}, even when accounting for differences in the IMF assumptions and when assuming that the bulge dynamical mass corresponds to the total stellar mass, the slope and normalisation of the $z=0$ relations differ \citep[see also discussion by][]{Kormendy13}.

\subsection{Outflow properties based on the integrated spectrum}\label{s:outflow}

Faint but clearly detected wings, different than the BLR, are seen in most of the strong emission lines in the integrated spectrum extracted from the central three by three spaxel of GS\_3073 (see Figure~\ref{f:fit}), which we interpret as tracing an outflow \citep[in line with other outflow indications by][]{Vanzella10, Grazian20}. The outflow emission appears almost symmetric, but slightly redshifted, suggesting that the majority of the outflow is pointing away from the observer \citep[see also][]{Vanzella10, Grazian20}. We measure the maximum projected outflow velocity as $v_{\rm out}=\langle v_{\rm outflow}\rangle+2\sigma_{\rm outflow}$, where $\sigma_{\rm outflow}$ is corrected for instrumental resolution \citep[e.g.][]{Genzel11, DaviesRL19}.
At the position of \oiii\ we measure $v_{\rm out}=1211\pm20$~km/s, while at the position of \ha\ we measure $v_{\rm out}=685\pm11$~km/s (see description of fitting model in Section~\ref{s:fitting}). At the \oiii\ position, we note an even more extended, faint redshifted wing in emission that is not captured by our best-fitting model (see left panel of Figure~\ref{f:fit}). The velocity separation of this emission reaches roughly 3100~km/s, suggesting a more complex and vigorous outflow.\footnote{
Another common definition of maximum projected outflow velocity, $v_{\rm out,2}=\langle v_{\rm outflow}\rangle+{\rm FWHM_{\rm outflow}}/2$, gives outflow velocities that are lower by about one third \citep[e.g.][]{Rupke05, Veilleux05, Arribas14}. In light of the high-velocity \oiii\ emission seen in the integrated spectrum, we continue our discussion of outflow properties with $v_{\rm out}$ as defined in the main text. See also \cite{FS19, DaviesRL20} for other definitions of $v_{\rm out}$.}

For the calculation of outflow properties such as the mass outflow rate $\dot{M}_{\rm out,ion}$ and mass loading factor $\eta_{\rm ion}=\dot{M}_{\rm out,ion}$/SFR, we use our measurements from the \ha\ outflow component, but note that they may correspond to lower limits given the higher velocity emission seen for the \oiii\ outflow. 
We adopt a simple model to estimate $\dot{M}_{\rm out,ion}$ from the \ha\ outflow component, assuming a photo-ionised, constant velocity ($v_{\rm out}$), spherical outflow of extent $R_{\rm out}$, following \cite{Genzel11, Newman12, FS19, DaviesRL19, DaviesRL20, Cresci23}:
\begin{multline}
\dot{M}_{\rm out,ion} = \frac{1.36m_{\rm H}}{\gamma_{\rm H\alpha}}\\ \times\frac{L_{\rm H\alpha,outflow}}{10^{42}{\rm erg\ s^{-1}}}\frac{1000{\rm cm^{-3}}}{n_{e,{\rm outflow}}}\frac{v_{\rm out}}{1000 {\rm km\ s^{-1}}}\frac{1 \rm kpc}{R_{\rm out}}\frac{M_\odot}{\rm yr},
\end{multline}
where $1.36m_{\rm H}$ is the effective nucleon mass for a 10 per cent helium fraction, $\gamma_{\rm H\alpha}=3.56\times10^{-25}$erg/cm$^{3}$/s is the \ha\ emissivity at $T=10^4 K$, $n_{e,{\rm outflow}}$ is the electron density in the outflow, and $L_{\rm H\alpha,outflow}$ is the \ha\ luminosity of the outflow  component.

The \sii\ lines are comparatively weak for GS\_3073, and no outflow line component in the [S\,{\sc{ii}}] doublet is preferred by our best fit. 
From the narrow component fit we infer $F_{\rm [S\,II]\lambda6716}/F_{\rm [S\,II]\lambda6731}=0.69\pm0.28$, corresponding to an electron density of $n_{e,\rm [S\,II]}\sim1869/{\rm cm^3}$ for $T=10^4$~K, using the calibration by \cite{Sanders16}. Alternatively, we can measure the electron density in the narrow and outflow component also from the \ariv$\lambda\lambda4711,4740$ ratio \citep[e.g.][]{Proxauf14}. However, in our case the \ariv$\lambda4711$ line is blended with \hei$\lambda4713$, making this measurement uncertain (see Figure~\ref{f:fit}). We cannot easily constrain the outflow component ratio, but report a density of $n_{e,\rm [Ar\,IV]}=3032/{\rm cm^3}$ based on the narrow component ratio, in line with this emission originating from higher density regions. 

For the density of the outflow component, since this is unconstrained by our data, we follow \cite{FS19} in assuming $n_{e,{\rm outflow}}=1000/{\rm cm^3}$ (their fiducial value for AGN-driven outflows; see also e.g.\ \citealp{Perna17, Kakkad18}).
With $L_{\rm H\alpha,outflow}=12.4\times10^{42}$~erg/s and adopting $R_{\rm out}=R_e\sim0.18$~kpc, we find $\dot{M}_{\rm out,ion}\sim98\pm2\,M_\odot$/yr. This is comparable to AGN with strong ionised outflows at $1<z<3$ \citep{FS19}. It is reasonable to assume that the total mass loss due to outflows is larger, since we are only probing the warm ionised gas phase with our measurements.
\cite{Rupke17} and \cite{Fluetsch19} have studied the relation between $M_{\rm BH}$ and $\dot{M}_{\rm out}$ from various gas phases in local galaxies. For GS\_3073 we find a mass outflow rate that is about twice as high as their best-fit relations, however well within the spread of individual measurements.

We can further estimate the mass loading factor $\eta_{\rm ion}=\dot{M}_{\rm out,ion}$/SFR. The literature SFR estimates from SED fitting for GS\_3073 vary in the range SFR$\sim30-410 M_\odot$/yr \citep{Barro19, Faisst20}. From the [C\,{\sc{ii}}]$\lambda158\micron$ luminosity, \cite{Barchiesi22} derive SFR$_{\rm [CII]}=16^{+14}_{-7}~M_\odot$/yr. 
Since the \ha\ narrow line flux likely has a strong AGN contribution, we instead use the SFR$_{\rm [CII]}$.\footnote{
However, we can use the narrow \ha\ flux to derive an upper limit of $60M_\odot$/yr on the SFR and a lower limit of $1.6$ on the ionised gas mass loading factor, consistent with the estimates from [C\,{\sc{ii}}].
}
From this we derive $\eta_{\rm ion}\sim6.1^{+5.0}_{-2.9}$.
This suggests the AGN in GS\_3073 is powerful enough to expel more mass from the galaxy than is currently consumed by star formation, in particular when considering the addition of cold and hot gas likely entrenched in the outflow.

We caution that the estimates of both $\dot{M}_{\rm out,ion}$ and $\eta_{\rm ion}$ are uncertain due to the substantial uncertainties regarding the ionised gas density and outflow geometry.

\begin{figure}
	\centering
	\includegraphics[width=\columnwidth]{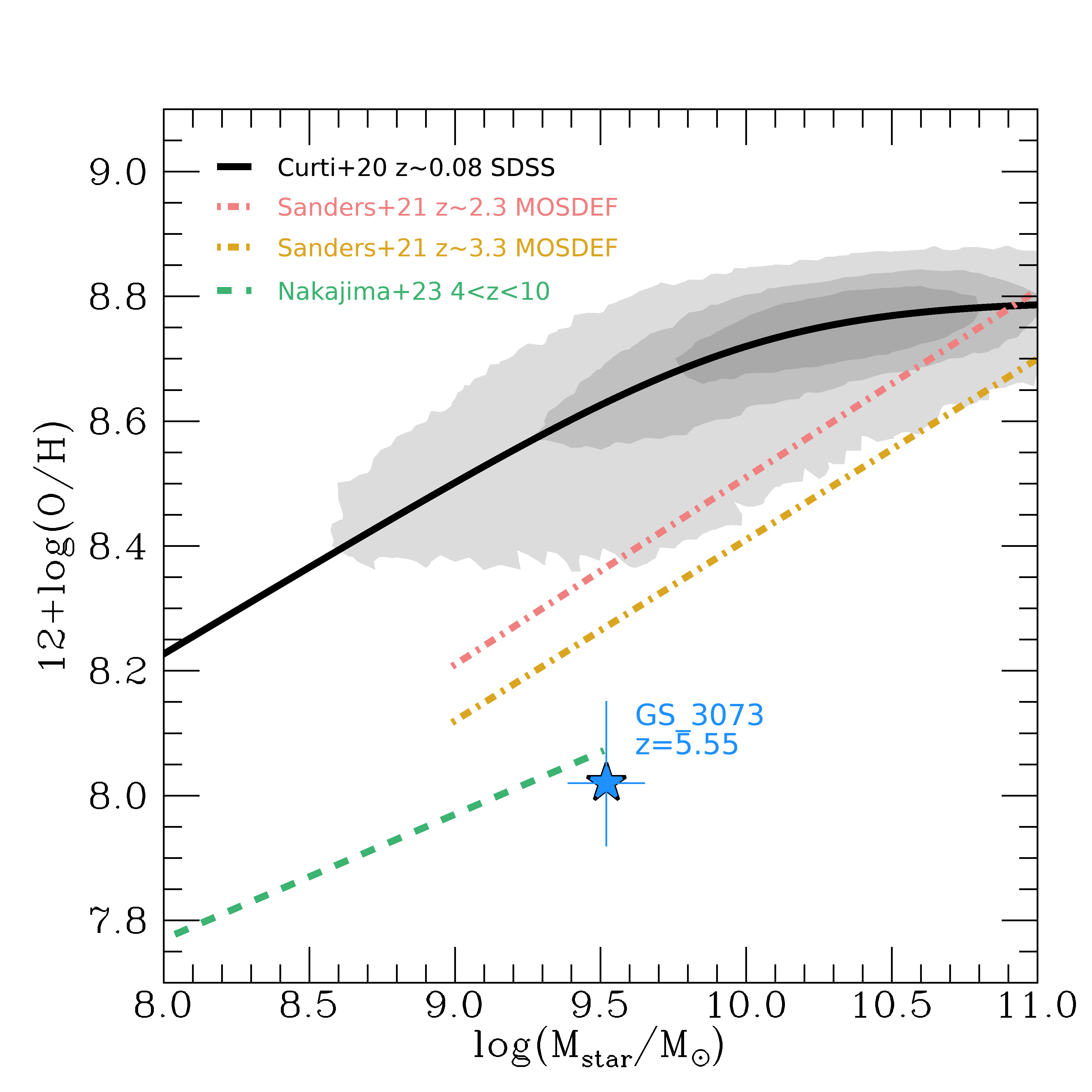}
	\caption{GS\_3073 in the mass-metallicity plane (blue star). The $z\sim0.08$ relation based on SDSS data by \citet{Curti20} is shown as grey shading with the best fit in black. The red and yellow dash-dotted lines show the best-fit relations at $z\sim2.3$ and $z\sim3.3$, respectively, by \citet{Sanders21} based on data from the MOSDEF survey \citep{Kriek15}. The dashed green line shows the best-fit relation obtained by \citet{Nakajima23} based on a compilation of early {\it JWST} data at redshifts $4<z<10$. GS\_3073 at $z=5.55$, although being slightly more massive, is compatible with the extrapolation of this relation.}
	\label{f:mzr}
\end{figure}

\subsection{Electron temperature and metallicity}\label{s:metallicity}

Our observations cover part of the auroral [O\,{\sc{iii}}]$\lambda4363$ emission line, which can be used in conjunction with [O\,{\sc{iii}}]$\lambda\lambda4959,5007$ to measure the electron temperature and gas phase metallicity (e.g.\ \citealp{Izotov06, Curti17}; and \citealp{Maiolino19} for a review). Although [O\,{\sc{iii}}]$\lambda4363$ is only partly covered in our R2700 data, we can still perform a simultaneous fit with the other emission lines in our spectrum by fixing the relative line position and the line widths to [O\,{\sc{iii}}]$\lambda5007$. While tentative, this gives an electron temperature of $T_e\approx14163^{+1339}_{-1439}$~K based on the narrow line ratio [O\,{\sc{iii}}]$\lambda4363$/[O\,{\sc{iii}}]$\lambda\lambda4959,5007$. For this calculation, we assume an electron density of $10^3/$cm$^3$, consistent within the uncertainties with our best-fit value based on the \sii\ doublet, and corresponding to the same value we adopt for the calculations of outflow properties (Section~\ref{s:outflow}). We note however that $T_{e,{\rm [OIII]}}$, i.e.\ the electron temperature of the high-ionisation zone, is relatively insensitive to the exact value of the electron density within a range of $10^2$-$10^4/$cm$^3$.
Following \cite{Dors2020}, assuming that the narrow line emission is dominated by AGN excitation, the measured $T_e$ corresponds to a metallicity of about $0.2~Z_\odot$, or
$12+\log$(O$^{++}$/H)$=8.00^{+0.12}_{-0.09}$.\footnote{We use solar metallicity $Z_\odot=0.014$ and $12+\log$(O/H)$=8.69$ \citep{Asplund09}.} By itself, this corresponds to a lower limit due to the unknown contribution of the other ionic species to the total oxygen abundance.

\begin{figure*}
	\centering
	\includegraphics[width=0.49\textwidth]{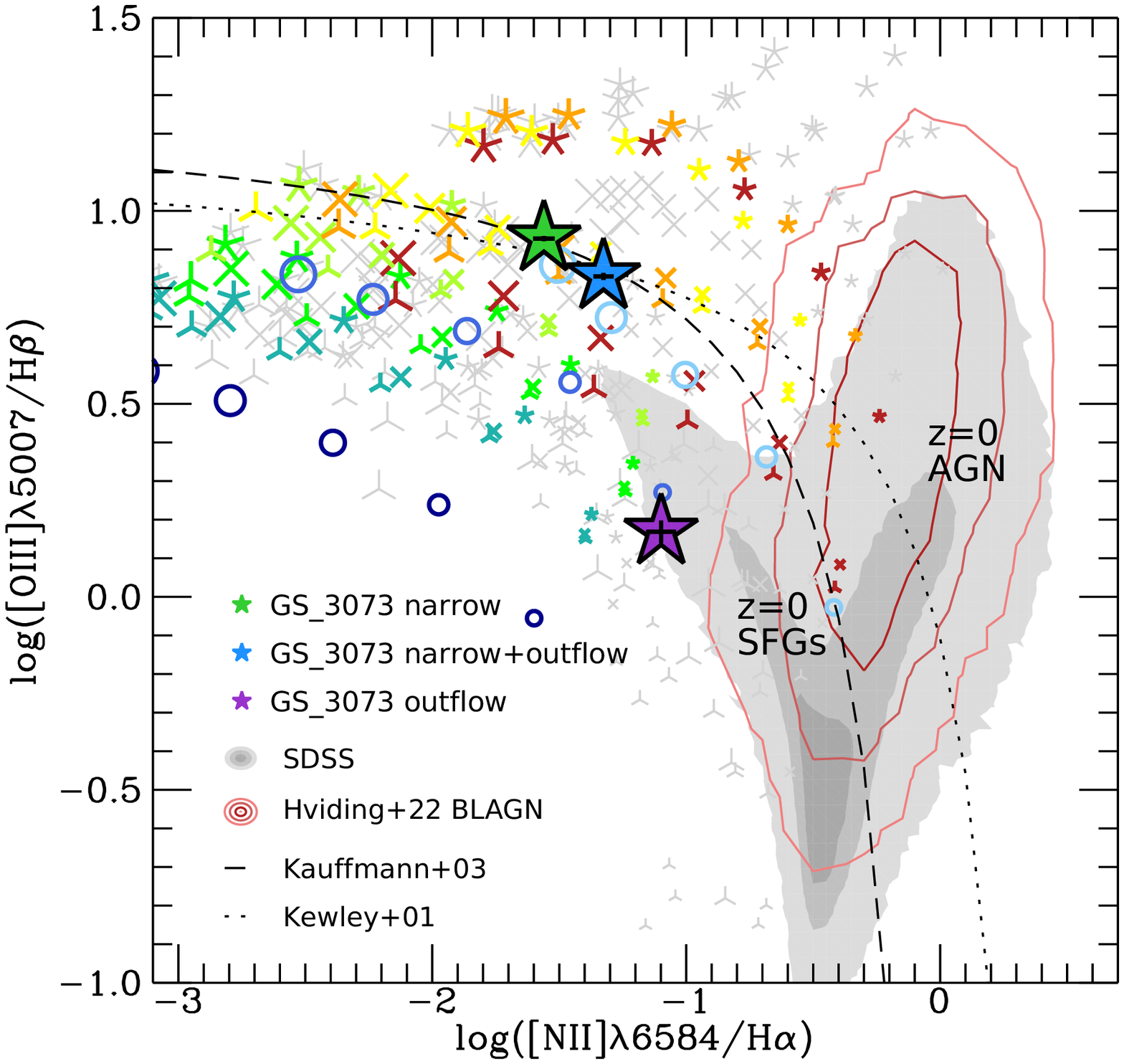}
	\includegraphics[width=0.49\textwidth]{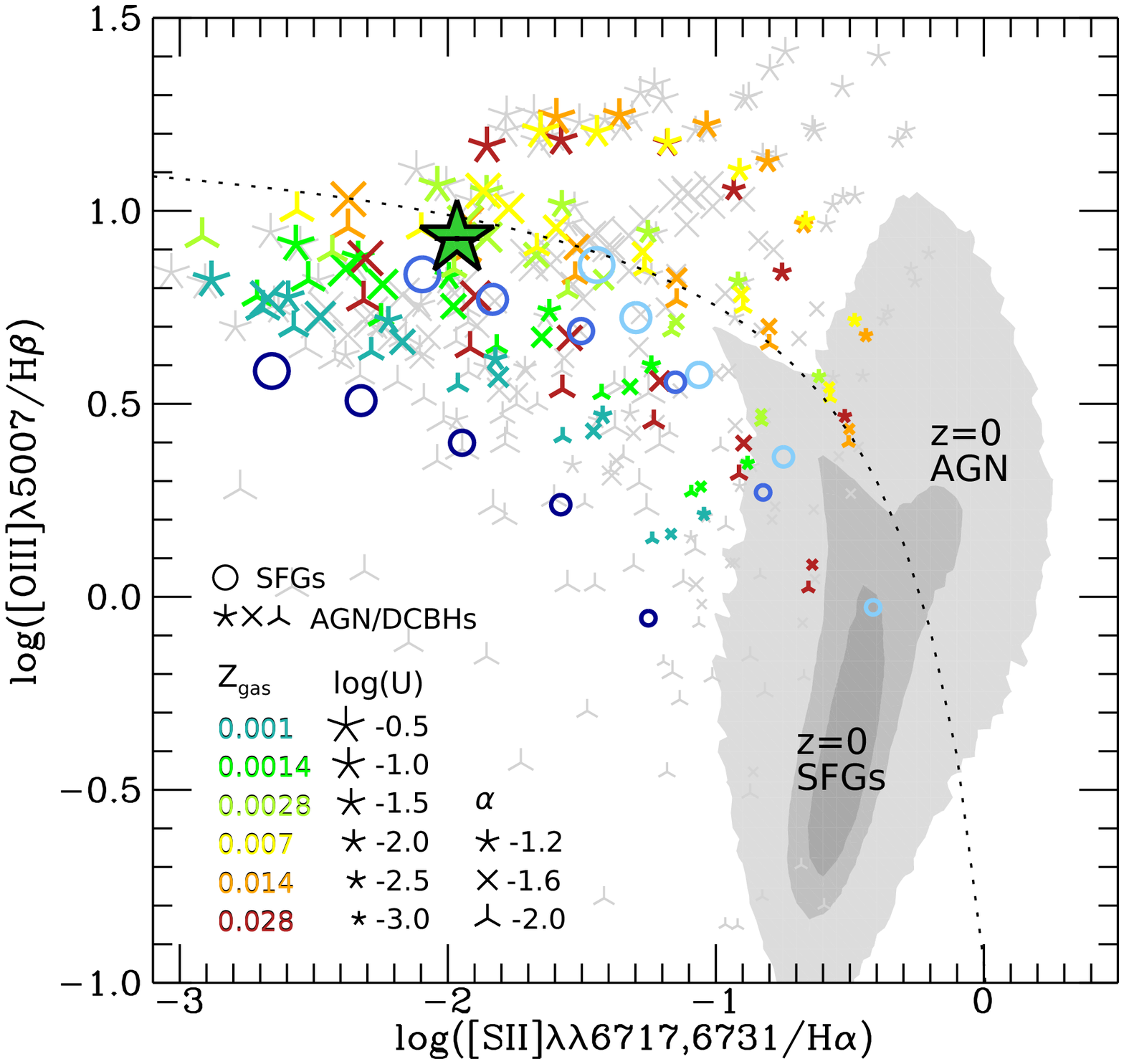}\caption{\oiii$\lambda5007$/\hb\ {\it vs. }\nii$\lambda6584$/\ha\ (left) and \oiii$\lambda5007$/\hb\ {\it vs. }\siid/\ha\ (right) diagnostic diagrams. The position of GS\_3073 is indicated as filled green (purple, blue) stars as constrained by its narrow (outflow, narrow+outflow) line ratios. Local galaxies from SDSS \citep{sdss7} are indicated by the grey shading with contours encompassing 98, 80, 50 per cent of the sample. The red contours encompass 95, 80, 50 per cent of SDSS broad line AGN as classified by \citet{Hviding22}. The dashed line indicates the demarcation by \citet{Kauffmann03} between galaxies primarily ionised by SF (left) and AGN (right). The dotted lines by \citet{Kewley01} include more extreme starbursts or composite objects among the star-forming galaxies to the left. The thin colored symbols (stars, crosses, triangles) show model grids for AGN and direct collapse black holes (DCBHs) by \citet{Nakajima22} for an accretion disc temperature of $T_{\rm bb}=2\times10^5$~K, power law indices to constrain the slope between the optical and $X-$ray bands ($\alpha=-1.2, -1.6, -2.0$, corresponding to stars, crosses, triangles), ionisation parameters ($-3.0<\log U<-0.5$ indicated by increasing symbol size), and gas phase metallicities ($Z_{\rm gas}=0.001-0.028$, teal to red colors). A gas density of $10^3/$cm$^3$ is assumed. Grey symbols show models with lower $T_{\rm bb}$. Open circles show model grids for galaxies using BPASS SEDs \citep{Eldridge17, Stanway18}, with binary evolution based on a \citet{Kroupa01} IMF with an upper mass cut of $300M_\odot$, and a stellar age of 10~Myr (see \citealp{Nakajima22} for details). Here, the dark to light colors indicate stellar metallicities in the range $Z_\star=0.0006-0.0084$. In these diagrams, GS\_3073 has line ratios compatible with being either a low-metallicity AGN or a low-metallicity SFG corresponding to these theoretical models, despite being located within the classical SFG regime.}
	\label{f:bpt}
\end{figure*}

However, in our R100 observations of GS\_3073 we cover the wavelength range $\lambda=0.6-5.3\micron$ which includes the [O\,{\sc{ii}}]$\lambda\lambda3727,3730$ doublet (see Appendix~\ref{a:r100_spec}). Together with our R2700 data,\footnote{We use the R2700 data to measure the line fluxes of [O\,{\sc{iii}}]$\lambda4363$ and [O\,{\sc{iii}}]$\lambda\lambda4959,5007$ because [O\,{\sc{iii}}]$\lambda4363$ is blended with H$\gamma$ in the lower-resolution data.} 
this allows us to constrain the full O/H abundance. We measure the [O\,{\sc{ii}}]$\lambda\lambda3727,3730$ flux from an aperture which gives [O\,{\sc{iii}}]$\lambda\lambda4959,5007$ flux consistent within 3 percent of our measurement from the R2700 data.
We find the contribution from O$^{+}$/H to be minor: exploiting the [O\,{\sc{ii}}]$\lambda\lambda3727,3730$ flux, and following the prescriptions from \cite{Dors2020} to account for the temperature of the low-ionisation region in Seyfert galaxies,
we derive a total oxygen abundance of $12+\log$(O/H)$=8.02^{+0.13}_{-0.10}$, corresponding to $0.21^{+0.08}_{-0.04}$ per cent solar metallicity.
We note that the contribution to the total abundance from even higher ionisation states of Oxygen is expected to be fully negligible, as the ionisation correction factor (ICF) computed on the basis of the He$^{+}$ and He$^{++}$ abundances is ICF(O)$=$N(He$^{+}$+He$^{++}$)/N(He$^{+}$)$=1.017$ \citep{Torres-Peimbert_1977,Izotov_1994}.
As shown in Figure~\ref{f:mzr}, this places our galaxy well below the gas-phase mass-metallicity relations measured from $z=0$ to $z\sim3$ \citep{Curti20a, Sanders21}, and in line with the relation inferred for star-forming galaxies at $4<z<10$ by \cite{Nakajima23}.

\section{Discussion}\label{s:discussion}

\subsection{An AGN in the SFG regime of the classical line ratio diagnostic diagrams}\label{s:bpt}

An important result of our work is that low metallicities in the early Universe blur the differences in classical diagnostic line ratios between galaxies that are primarily ionised by SF {\it vs.\ }AGN activity.
This can be appreciated in Figure~\ref{f:bpt}, where we show the placement of our galaxy in the line ratio diagnostic diagrams \oiii/\hb\ {\it vs.} \nii/\ha\ (left; \citealp{Baldwin81}) and \oiii/\hb\ {\it vs.} \siid/\ha\ (right; \citealp{Veilleux87}). 
Our source (filled green, purple, and blue stars for narrow, outflow, and narrow+outflow line ratios, respectively) has an \nii/\ha\ ratio which is lower than for the majority of local SFGs, and well separated from the local AGN branch. The \nii/\ha\ and \sii/\ha\ narrow line ratios of GS\_3073 are also much lower than those of massive SFGs at $1<z<3$, while the \oiii/\hb\ ratio is above average (\citealp[e.g.][]{Steidel14, Strom17, Curti20, Topping20}; see \citealp{Maiolino19} for a review). Despite being an AGN, the low \nii/\ha\ and \sii/\ha\ line ratios place our source into the star-forming regime of the classical $z=0$ line ratio diagnostic diagrams.

Theoretical models do predict that AGN in the early Universe might populate the `star-forming' regime of the classical line ratio diagnostics, to the left of the $z=0$ AGN branch, mainly due to their lower metallicities. This is shown in Figure~\ref{f:bpt} through thin stars, crosses, and triangles which represent model predictions by \cite{Nakajima22} for AGNs with varying metallicities, ionisation parameters, and power law indices (see also e.g.\ \citealp{Groves06, Kewley13, Feltre16, Gutkin16, Hirschmann17, Hirschmann19}; and see \citealp{Hirschmann22} for a study post-processing cosmological simulations). Models with an accretion disc temperature of $T_{\rm bb}=2\times10^5$~K are shown in color, while models with $T_{\rm bb}=5\times10^4$~K and $T_{\rm bb}=1\times10^5$~K are shown as grey symbols.

However, if we consider theoretical model predictions for early galaxies with massive stars, indicated by blue/purple circles in Figure~\ref{f:bpt}, they are also in agreement with the location of GS\_3073. These models suggest that the classical line ratio diagnostic diagrams alone cannot be used to distinguish SFGs and AGN in the early Universe.
Our data demonstrate that high$-z$ AGN can indeed populate the `star-forming' regime of the classical line ratio diagnostics \citep[see also][]{Kocevski23}.

\begin{figure*}
	\centering
	\includegraphics[width=0.49\textwidth]{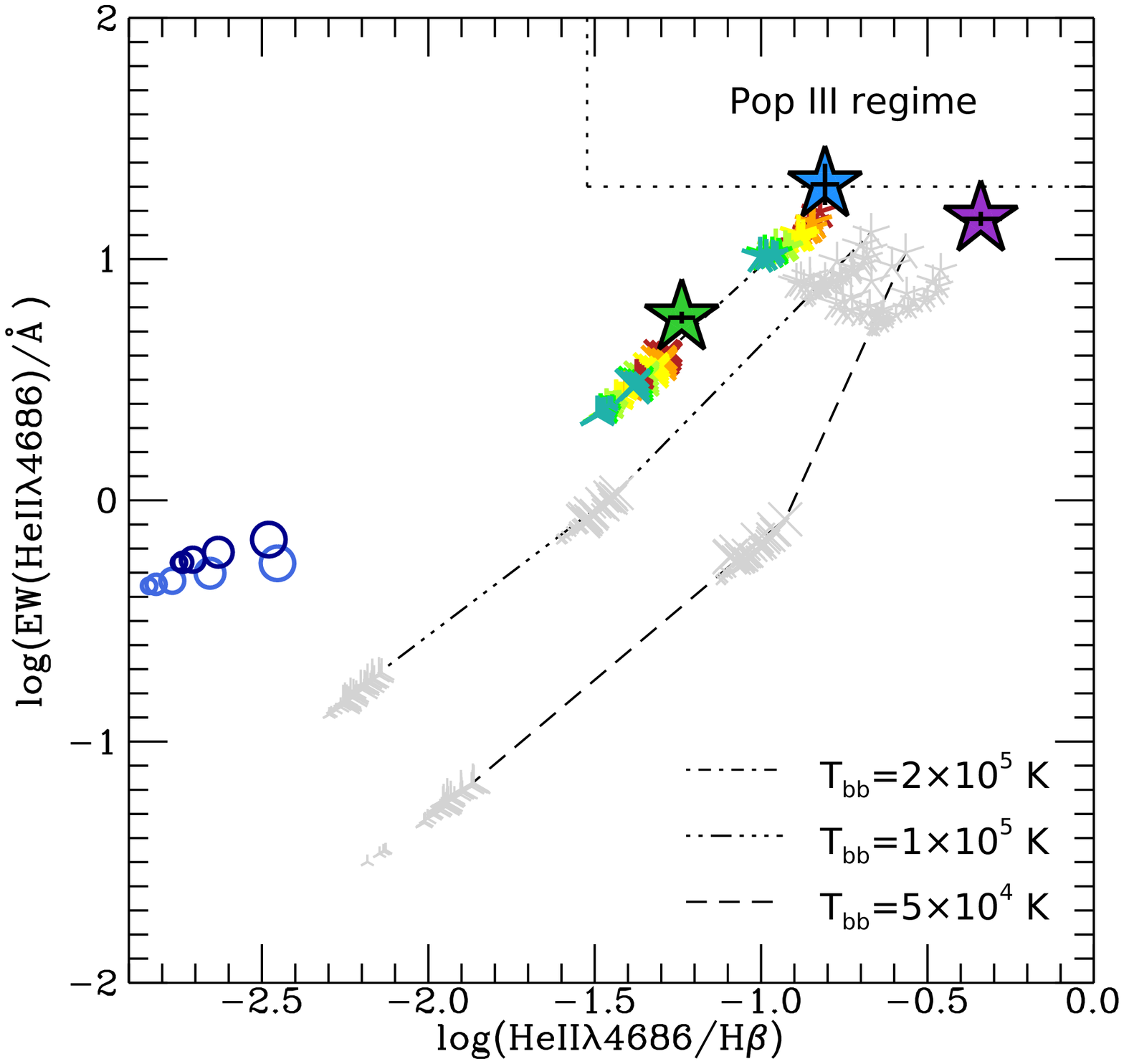}
	\includegraphics[width=0.49\textwidth]{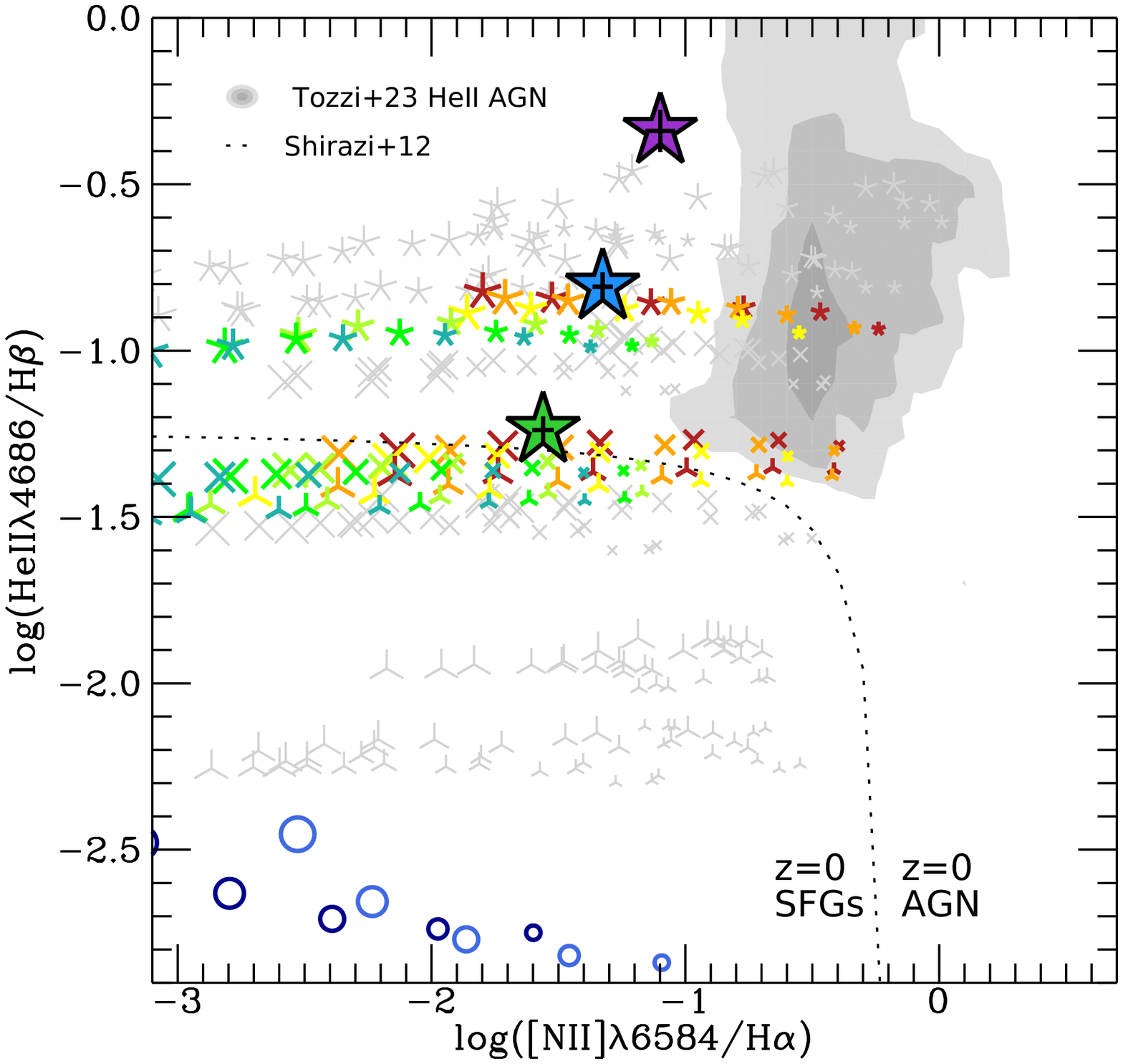}
	\caption{EW(\heii) {\it vs. }\heii/\hb (left) and \heii/\hb\ {\it vs. }\nii$\lambda6584$/\ha\ (right) diagnostic diagrams. Symbols are the same as in Figure~\ref{f:bpt}: thin stars, crosses and triangles show evolved AGN and DCBHs, circles show evolved SFGs, and the position of GS\_3073 is indicated by the large filled stars (green: narrow, blue: narrow+outflow, purple: outflow). In the left panel, lines of constant $T_{\rm bb}=5\times10^4$~K, $1\times10^5$~K, $2\times10^5$~K corresponding to dashed, dash-triple-dotted, and dash-dotted lines. In the left panel, the dotted region in the top right indicates the regime expected for Pop\,{\sc{iii}} stars. In the right panel, narrow-line ratios from local He\,{\sc{ii}}-selected AGN by \cite{Tozzi23} are indicated by the grey shading with contours encompassing 94, 80, 50 per cent of the sample. The dotted line indicates the demarcation by \cite{Shirazi12} separating SFGs (left) from AGN (right) at $z=0$.
    In the EW(\heii) {\it vs. }\heii/\hb\ diagram, GS\_3073 is compatible with an AGN with a hot accretion disc, and clearly separated from SFGs. This is also the case for the \heii/\hb\ {\it vs. }\nii$\lambda6584$/\ha\ diagnostic (right), although the locally constrained demarcation line by \citet{Shirazi12} appears to evolve with redshift, if we consider the location of model predictions particularly for AGNs with $\alpha=-2.0$ (thin triangles).}
	\label{f:heii}
\end{figure*}

We note that also some low-metallicity, low$-z$ AGN are found in the star-forming region of the classical line ratio diagnostic diagrams, although typically with higher \nii/\ha\ ratios ($>0.1$) than our source \citep[but see also e.g.][]{Simmonds16, Cann20, Burke21}. This can be appreciated through the location of $z\lesssim0.3$ AGN with broad Balmer lines by \cite{Hviding22}, indicated by red contours in the left panel of Figure~\ref{f:bpt} \citep[see also e.g.][]{Shirazi12, Kawasaki17, Keel19}.

Since we know that GS\_3073 is an AGN, and we have an estimate of its gas phase metallicity from the narrow line ratios (Section~\ref{s:metallicity}), we can use this information together with the theoretical model predictions to further constrain the ionisation parameter, $U$, and the power law index of the energy slope between the optical and $X-$ray bands, $\alpha$. Based on our measurement of $Z_{\rm gas}/Z_\odot=0.21$, we focus on models with $Z_{\rm gas}=0.0028$ (20 per cent solar; yellow-green symbols by \citealp{Nakajima22} in Figure~\ref{f:bpt}) in the BPT diagram (left panel), in which our measured line ratios have the highest $S/N$. We find that our BPT narrow line ratios are consistent with $\log(U)=-2.0$ and $\alpha=-1.2$.

\subsection{Using He\,{\sc{ii}}$\lambda4686$ to discriminate SFGs from AGN}\label{s:heii}

To account for the unique conditions in the early Universe, alternative diagnostic diagrams have been proposed to separate SFGs from AGN. Several of them rely on the properties of the \heii\ emission we also detect in our galaxy \citep[e.g.][]{Shirazi12, Baer17, Nakajima22}.\footnote{
\heii\ emission can also be associated with Wolf-Rayet stars, however in this case is blended with lines such as N\,{\sc{iii}}$\lambda4640$ to form the so-called `blue bump' at $\lambda=4600-4680$\AA, which is not seen in GS\_3073 \citep[e.g.][]{Brinchmann08}. Other sources of \heii\ have been proposed including $X-$ray binaries and fast shocks to explain observations in some low-redshift, low-metallicity star-forming dwarf galaxies \citep[e.g.][]{Thuan05, Kehrig15, Schaerer19, Umeda22}. However, for GS\_3073 the AGN nature of the ionising radiation is unambiguous through the detection of the BLR.}

In Figure~\ref{f:heii} we show the placement of our source in two diagnostic diagrams utilising \heii. In the left panel, we show the equivalent width EW(\heii) as a function of \heii/\hb. This diagram provides constraints on the shape of the ionising spectrum and the temperature of the accretion disc: the presence of \heii\ with an ionisation potential of 54.4~eV requires sources of hard ionising radiation. Its equivalent width increases with increasing fraction of highly ionising photons over non-ionising photons, and has therefore been promoted as an indicator of Population {\sc{iii}} stars (their proposed location is indicated by the dotted grey box). 
\heii/\hb\ effectively constrains the shape of the ionising spectrum through the ratio of ionising photons with $E>54.4$~eV to $E>13.6$~eV. 

The location of GS\_3073 (filled green, purple, blue stars for the narrow, outflow, narrow+outflow line ratios) in this diagram is better reproduced by models using a high accretion disc temperature of $T_{\rm bb}=2\times10^5$~K (symbols connected by the dash-dotted line), possibly indicating an even higher temperature. A high temperature of the accretion disc is generally associated with smaller black holes, and in line with the fact that the black hole in this AGN is smaller than most black holes inferred for more luminous quasars at similar redshift. Both the high equivalent width of \heii\ and the high \heii/\hb\ ratio clearly separate our galaxy from model predictions of galaxies without an AGN (blue/purple circles).

The situation is similar for the line ratio diagnostic in the right panel of Figure~\ref{f:heii}, originally suggested by \cite{Shirazi12} to distinguish more clearly sources primarily ionised by SF {\it vs. }AGN in the local Universe. 
Here we also show as grey shading $z=0$ data by \cite{Tozzi23}, who selected AGN-dominated spaxels (with $>50\%$ of \heii\ flux excited by the AGN) from MaNGA galaxies \citep{Bundy15}. Note that the authors subtract any BLR emission before measuring the line ratios. Similar to the BPT diagram, local \heii-selected AGN have higher \nii/\ha\ ratios compared to GS\_3073.
Considering the \cite{Nakajima22} model predictions for early SFGs (circles) and AGN (other small colored symbols), the separation between these still prevails at high redshift, even though the demarcation line might evolve over time as suggested by the placement of the model predictions compared to the $z=0$ data.

While the classical line ratio diagnostic diagrams cannot help to identify the primary ionisation source for low-metallicity galaxies (see Section~\ref{s:bpt}), the placement of GS\_3073 in the \heii\ diagnostics discussed here with respect to theoretical predictions suggests that those can be used instead.

\begin{figure*}
	\centering
	\includegraphics[width=0.49\textwidth]{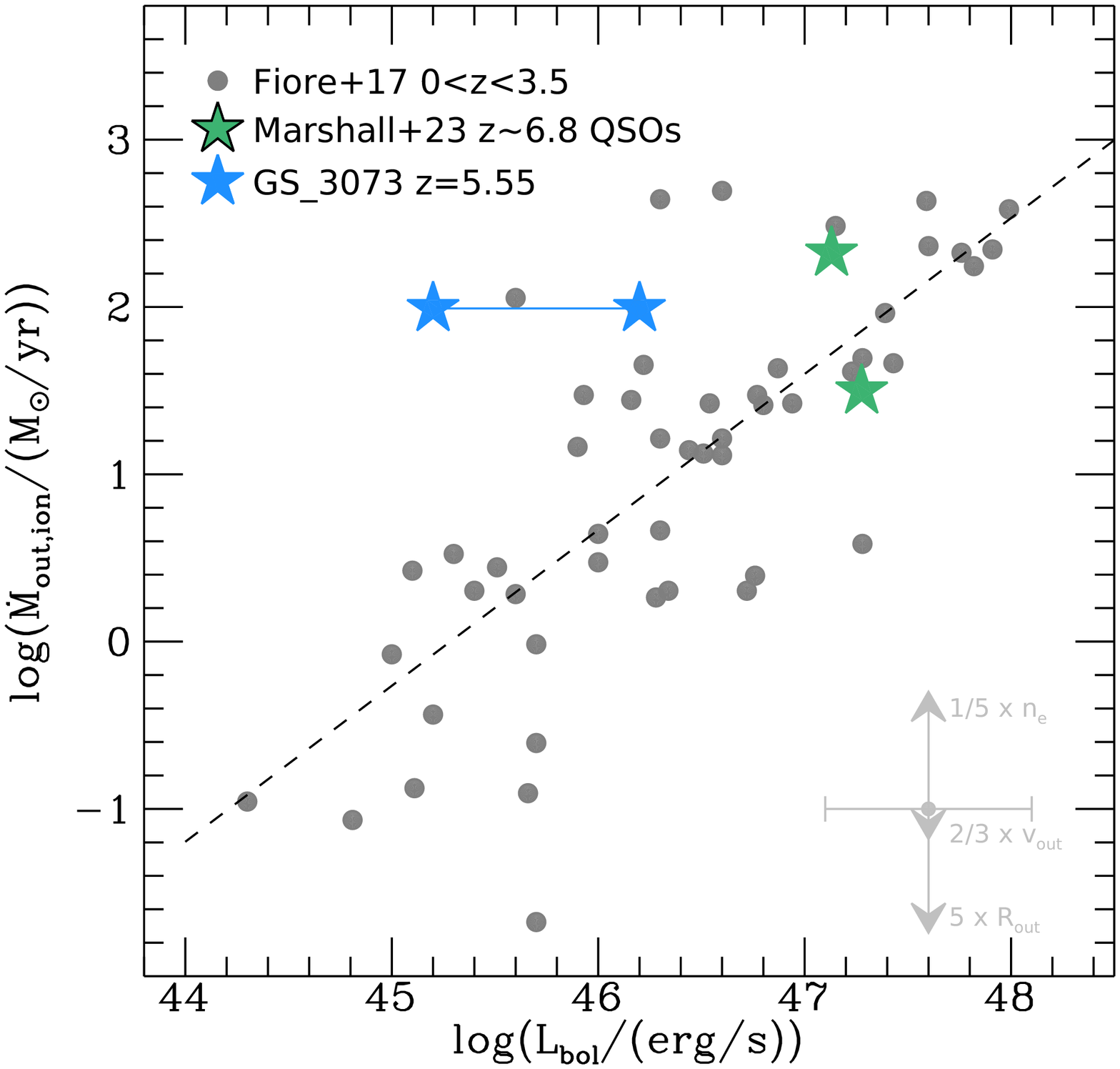}
	\includegraphics[width=0.49\textwidth]{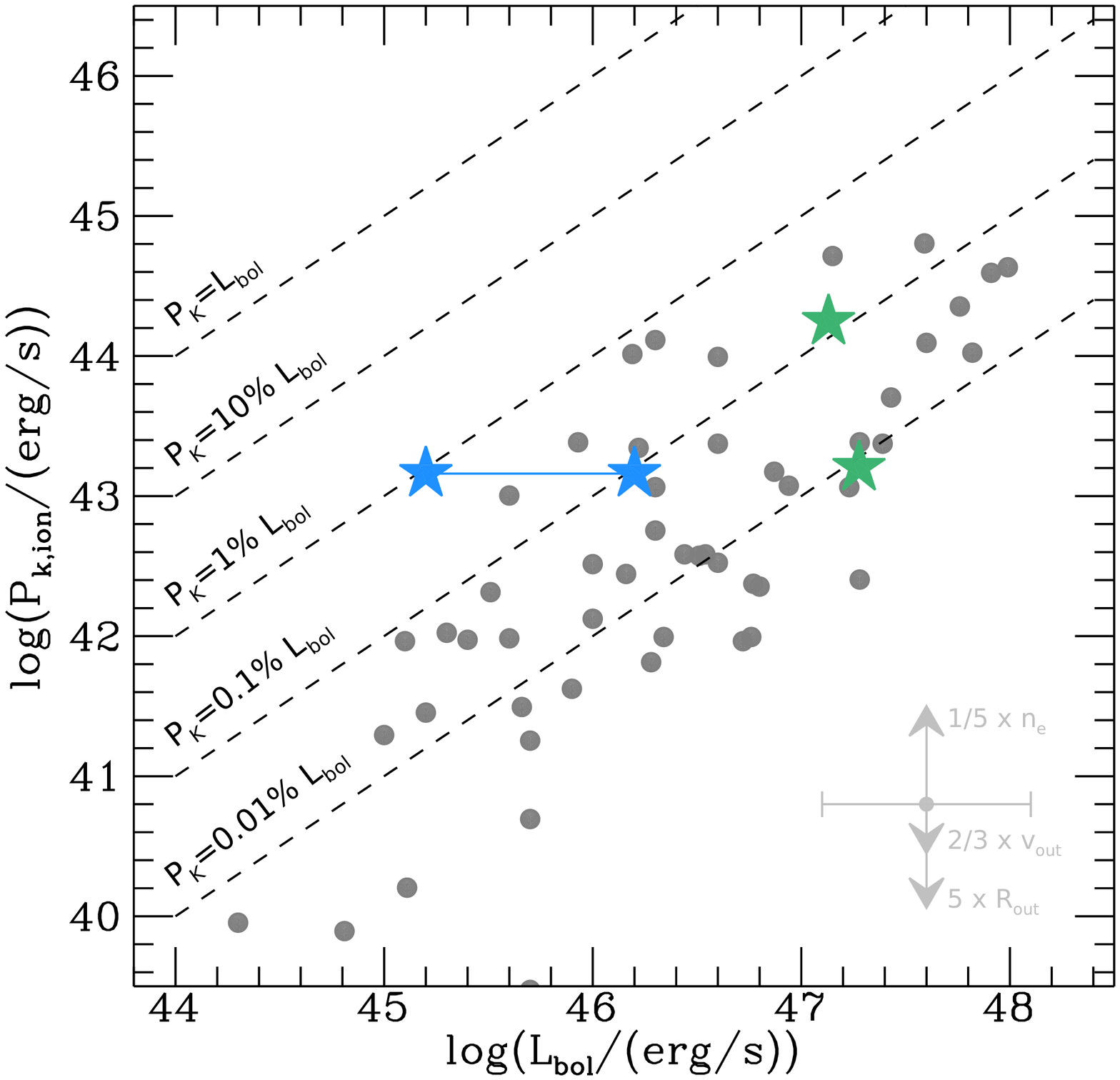}
	\caption{Mass outflow rate $\dot{M}_{\rm out,ion}$ (left) and kinetic power $P_{k,\rm ion}$ (right) of the warm ionised gas phase as a function of AGN bolometric luminosity $L_{\rm bol}$. Grey circles show ionised gas outflows at $0<z<3.5$ by \citet{Fiore17}, the green stars show two $z\sim 6.8$ QSOs from the GA-IFS GTO program by \cite{Marshall23}, and the blue stars show the range of bolometric luminosities inferred for GS\_3073 as discussed in Section~\ref{s:bhmass}. Here, we have scaled the measurements by \cite{Fiore17} and \cite{Marshall23} to correspond to an electron density in the outflow of $n_e=1000$/cm$^{3}$, as assumed in this work. 
    For $L_{\rm bol}$ we indicate a representative error bar of 0.5~dex, and for $\dot{M}_{\rm out,ion}$ and $P_{k,\rm ion}$ we indicate how much they would change if $n_e$, $R_{\rm out}$, or $v_{\rm out}$ would be different by factors $1/5$, $5$, and $2/3$, respectively (see main text for details).
    The dashed line in the left panel is a linear fit in logarithmic scales to the data by \citet{Fiore17}. The dashed lines in the right panel show constant ratios of $P_k$ and $L_{\rm bol}$. The outflow energetics of GS\_3073 in relation to the bolometric AGN luminosity are consistent with lower$-z$ scaling relations, suggesting that the driving mechanisms in $z>5$ AGN do not differ from their lower$-z$ counterparts.}
    \label{f:mout}
 \end{figure*}

\subsection{Massive black holes in the early Universe}

Our measurement of $\log(M_{\rm BH}/M_\odot)\sim8.2$ from the \ha\ broad line region is among the few measured `lower mass' black holes at $z>5$ \citep[see also][]{Kocevski23}. Still, the black hole of GS\_3073 appears overly massive compared to local scaling relations by \cite{Reines15} and \cite{Kormendy13}, and compared to theoretical model predictions \citep[e.g.][based on the $2\micron$ flux; priv. comm.]{Trinca23}, similar to other black hole mass measurements at higher redshift. 

If black holes at higher redshift are comparatively more massive, this could suggest a more rapid growth of black holes in the early Universe, fuelled by larger gas fractions, or more efficient accretion.
Higher accretion rates could plausibly be achieved in systems with a higher density. This idea is encapsulated in the analytical model by \cite{Chen20}, where at fixed stellar mass smaller SFGs host more massive black holes due to higher central densities. Assuming our fiducial $R_e=0.18$~kpc and $\log(M_\star/M_\odot)=9.52$, their $M_{\rm BH}-R_e-M_{\rm star}$ relation (their equation C12) predicts a black hole mass of $\log(M_{\rm BH}/M_\odot)\sim7.0$. This prediction is above the local $M_{\rm BH}-M_{\rm star}$ relation (see Figure~\ref{f:mbh}) in line with our findings, but still lower than our measurement of $\log(M_{\rm BH}/M_\odot)\sim8.2$.

Interestingly, overmassive black holes are also found in a comparable region of the $M_{\rm BH}-M_\star$ parameter space in some $z\lesssim1$ dwarf galaxy AGN \citep[e.g.][]{Burke22, Mezcua23, Siudek23}. These galaxies may evolve onto the local $M_{\rm bulge}-M_{\rm BH}$ relation by $z=0$ \citep{Mezcua23}.
Larger black hole masses at earlier times are also predicted by some cosmological simulations, while others show the opposite trend \citep[see][]{Habouzit21, Habouzit22}. Consolidating a picture of high$-z$ black holes masses in relation to their host galaxy properties over a wide range in masses could therefore serve as a powerful discriminant of feedback implementations.

\subsection{AGN feedback and enrichment of the intergalactic medium}

Theoretical work suggests that AGN feedback is crucial in quenching galaxies with host galaxy masses close to the Schechter mass \citep[$\log(M_{\rm star}/M_\odot)\sim11$; e.g.][]{diMatteo05, Croton06, Bower06, Hopkins06, Cattaneo06, Somerville08}, and this is supported by observational evidence \citep[e.g.][]{Veilleux05, McNamara07, Fabian12, Genzel14, Harrison14, Harrison16, Heckman14, FS14, FS19}. Its impact has also been demonstrated empirically \citep[e.g.][]{Penny18, ManzanoKing19, Mezcua19, Liu20, Davis22} and theoretically \citep[e.g.][]{Koudmani19, Koudmani21} for galaxies with much lower masses. 

In Figure~\ref{f:mout} we compare the mass outflow rate (left) and kinetic power $P_k=0.5\dot{M}_{\rm out}v_{\rm out}^2$ (right) measured from the fit to the spectrum of GS\_3073 integrated over the central three by three spaxels (Figure~\ref{f:fit}, Section~\ref{s:outflow}) as a function of AGN bolometric luminosity to local and lower redshift sources ($z<3.5$) by \cite{Fiore17}, and to two $z\sim6.8$ QSOs by \cite{Marshall23}. The outflow energetics of GS\_3073 are consistent with the scalings derived from the lower$-z$ data, suggesting that the driving mechanisms in $z>5$ AGN do not differ strongly from their lower$-z$ counterparts. Interestingly, the kinetic power is only $\sim0.1-1.0$\% of the radiative luminosity of the AGN. This is generally considered as an indication of the outflow being little effective in depositing energy into the ISM and therefore not providing major feedback onto the galaxy, at least not in the direct ejective mode. As predicted by many theoretical models \citep[e.g.][]{Gabor14,Roos15,Bower17,NelsonD19,Zinger20}, while the ionised outflow may have little impact on the ISM (possibly because of poor coupling), it can potentially escape into the circum-galactic medium (CGM). Thus it may contribute to its heating, hence suppressing fresh gas accretion and possibly resulting in delayed feedback in the form of starvation. 

We emphasize that the quantities discussed here and shown in Figure~\ref{f:mout} are subject to substantial uncertainties. As mentioned in Section~\ref{s:outflow}, the unknown electron density of the outflowing material and the uncertain outflow geometry hamper a robust estimate of $\dot{M}_{\rm out,ion}$ and $P_{k,\rm ion}$. As a reference, for $\dot{M}_{\rm out,ion}$ and $P_{k,\rm ion}$ we indicate in Figure~\ref{f:mout} by arrows how much they would change if $n_e$, $R_{\rm out}$, or $v_{\rm out}$ would be different by factors $1/5$, $5$, and $2/3$, respectively. Changes in these quantities, e.g.\ because of a different definition of the outflow velocity, or a different assumption on the gas density in or the extent of the outflow dominate the uncertainties (see Section~\ref{s:outflow}). For the uncertainty on $L_{\rm bol}$, we indicate $0.5$~dex, motivated by the range of values derived for GS\_3073, and the estimate by \cite{Fiore17}.

We measure relatively large projected outflow velocities and estimate a high mass loading factor for GS\_3073 based on our fit to the integrated spectrum over the central three by three spaxels (Figure~\ref{f:fit}, Section~\ref{s:outflow}). At the same time, we infer a relatively low dynamical mass for our galaxy (Section~\ref{s:mdyn}). This suggests that indeed a substantial fraction of the material expelled by the outflow could escape the potential well of the galaxy to enrich the CGM and even inter-galactic medium (IGM).

To quantify this, we estimate the escape velocity at radius $r$ assuming an isothermal sphere following \cite{Arribas14} as 
\begin{equation}\label{eq:vesc}
v_{\rm esc}\approx \sqrt{\frac{2M_{\rm dyn}G [1+\ln(r_{\rm max}/r)]}{3r}},
\end{equation}
where $r_{\rm max}$ is the truncation or halo radius. Evaluating at $r=R_e=0.18$~kpc, with $r_{\rm max}=100r$, we find $v_{\rm esc}\sim390$~km/s. Comparing this to our measured outflow velocity of $v_{\rm out}=685$~km/s, with some material reaching velocities of few 1000~km/s as based on the outflow [O{\sc{iii}}] component, this indicates that indeed a sizable fraction of the outflowing gas could escape the galaxy's potential well. 
From the distribution of outflow velocities in our best-fit outflow components, we find that about $5-20$ per cent of the emitting ionised gas has velocities in excess of $v_{\rm esc}\sim390$~km/s, when considering both the components around H$\alpha$ and around [O{\sc{iii}}]. 
This suggests that at least $6-20 M_\odot$/yr of warm ionised gas could escape the galaxy potential if feedback would sustain the measured outflow velocities and mass outflow rates.
This material could thus contribute to metal enrichment of the IGM at this early time in cosmic history.

We note that this estimate likely represents a lower limit for two reasons: firstly, the measured outflow velocities are projected, i.e.\ intrinsic outflow velocities will be even larger. Secondly, we do not account for gas phases other than warm ionised gas plausibly entrenched in the outflow; in particular we expect contributions from cold (molecular and neutral) gas, that are likely dominating the outflow mass budget \citep[see e.g.][]{Rupke13, HerreraCamus19, RobertsBorsani20, Fluetsch21, Avery22, Baron22, Cresci23}.

\section{Conclusions and outlook}\label{s:conclusions}

We have presented deep {\it JWST}/NIRSpec integral field spectroscopy of the galaxy GS\_3073 at $z=5.55$. We have focused on the high spectral resolution spectrum (G395H, $R\sim2700$) obtained with 5~h on-source, while we have also used a shorter exposure (1~h) prism spectrum ($R\sim100$). The high resolution spectrum has revealed about 20 rest-frame optical nebular emission lines, some of which are detected with very high $S/N$, and another 14 lines/doublets are visible in the prism spectrum. The main results of our analysis are:

\begin{itemize}
    \item Permitted lines, such as H$\alpha$, H$\beta$, He\,{\sc{i}} and He\,{\sc{ii}}, are characterized by a broad component (not observed in the forbidden lines) which can be unambiguously interpreted as tracing the Broad Line Region (BLR) around an accreting supermassive black hole, and clearly identifying this as a (type 1.8) AGN.

    \item From the narrow line ratios, we measure a gas phase metallicity of $Z_{\rm gas}/Z_\odot\sim0.21$, lower than what has been inferred for both more luminous AGN at similar redshift and lower$-z$ AGN.
    
    \item We empirically show that classical line ratio diagnostics \citep{Baldwin81, Veilleux87} cannot be used to distinguish between the primary ionisation source (AGN or SF) for such low-metallicity systems, whereas different diagnostic diagrams involving He\,{\sc{ii}}$\lambda4686$ prove useful.
    
    \item We measure the central black hole mass of GS\_3073 to be $\log(M_{\rm BH}/M_\odot)\sim8.2$. While this places our galaxy at the lower end of known high$-z$ black hole masses, it still appears to be over-massive compared to its host galaxy properties such as stellar mass or dynamical mass.
    
    \item We detect an outflow with velocity $v_{\rm out}=685$~km/s and a mass outflow rate of about $100 M_\odot/$yr, suggesting that GS\_3073 is able to enrich the intergalactic medium with metals one billion years after the Big Bang.
    
\end{itemize}

Additional JWST data, especially spectroscopic surveys with the MSA, will certainly discover more AGN like the one presented in this paper and will allow an assessment of the AGN and black hole census in the early Universe. It will also be possible to study the impact of early AGN feedback on the first massive galaxies, especially with IFS follow-up observations.

Our paper has highlighted that the detection of AGN cannot rely entirely on the classical diagnostics diagrams that have been developed and used locally and at intermediate redshifts ($z\sim1-3$). Other techniques have to be adopted, and the detection of broad components of the permitted lines (not accompanied by similar components on the forbidden lines) provide a clear and unambiguous way to identify accreting black holes. We note that this method requires a spectral resolution of at least $R>500$ in order to properly identify and disentangle broad and narrow components. Within this context, NIRSpec's Prism is borderline for this methodology, even in its reddest part, where its spectral resolution reaches $R\sim300$. The medium-resolution gratings are optimally suited for the detection of broad components. The high-resolution gratings, such as the one adopted in this paper, are excellent for the detailed modelling of the line profile when the $S/N$ is very high, but it may miss broad wings in the noise in the case of lower-$S/N$ spectra.

\section*{Acknowledgements}

We are grateful to the anonymous referee for a constructive report that helped to improve the quality of this manuscript.
We thank Takuma Izumi for sharing their compilation of black hole and dynamical masses of $z\gtrsim6$ QSOs published by \cite{Izumi19, Izumi21}. 
We thank Kimihiko Nakajima for providing the theoretical model grids published by \cite{Nakajima22}.
We thank Raphael Erik Hviding for sharing BPT line ratios of their local broad line AGN sample published by \cite{Hviding22}.
We thank Giulia Tozzi for sharing emission line ratios of their local He\,{\sc{ii}}-selected AGN published by \cite{Tozzi23}.
We acknowledge the JADES team for prompting a closer investigation of the source morphology in our R100 NIRSpec-IFS data.
We are grateful to Raffaella Schneider, Alessandro Trinca, Giulia Tozzi, and Stijn Wuyts for discussing various aspects of this work, and to Sandy Faber, Rachel Bezanson, and William Keel for valuable input.
We thank Taro Shimizu, Mar Mezcua, and Masafusa Onoue for helpful comments on an earlier version of this manuscript.
AB, GCJ acknowledge funding from the ``FirstGalaxie'' Advanced Grant from the European Research Council (ERC) under the European Union's Horizon 2020 research and innovation programme (Grant agreement No.~789056).
FDE, RM, JS acknowledge support by the Science and Technology Facilities Council (STFC), from the ERC Advanced Grant 695671 ``QUENCH''.
BRP, MP, SA acknowledge support from the research project PID2021-127718NB-I00 of the Spanish Ministry of Science and Innovation/State Agency of Research (MICIN/AEI).
GC acknowledges the support of the INAF Large Grant 2022 ``The metal circle: a new sharp view of the baryon cycle up to Cosmic Dawn with the latest generation IFU facilities''.
H{\"U} gratefully acknowledges support by the Isaac Newton Trust and by the Kavli Foundation through a Newton-Kavli Junior Fellowship. 
MAM acknowledges the support of a National Research Council of Canada Plaskett Fellowship, and the Australian Research Council Centre of Excellence for All Sky Astrophysics in 3 Dimensions (ASTRO 3D), through project number CE170100013.
MP acknowledges support from the Programa Atracci\'on de Talento de la Comunidad de Madrid via grant 2018-T2/TIC-11715.
PGP-G acknowledges support  from  Spanish  Ministerio  de  Ciencia e Innovaci\'on MCIN/AEI/10.13039/501100011033 through grant PGC2018-093499-B-I00.
RM acknowledges funding from a research professorship from the Royal Society.
S.C acknowledges support from the European Union (ERC, WINGS, 101040227).
The Cosmic Dawn Center (DAWN) is funded by the Danish National Research Foundation under grant no.140.
This work has made use of the Rainbow Cosmological Surveys Database, which is operated by the Centro de Astrobiología (CAB), CSIC-INTA, partnered with the University of California Observatories at Santa Cruz (UCO/Lick, UCSC).

%%%%%%%%%%%%%%%%%%%%%%%%%%%%%%%%%%%%%%%%%%%%%%%%%%
%\section*{Data Availability}



\bibliographystyle{aa}
\bibliography{literature} % if your bibtex file is called example.bib

%%%%%%%%%%%%%%%%%%%%%%%%%%%%%%%%%%%%%%%%%%%%%%%%%%

%%%%%%%%%%%%%%%%% APPENDICES %%%%%%%%%%%%%%%%%%%%%

\appendix

\section{Line fluxes of He\,{\sc{i}}}\label{a:hei}

In Table~\ref{t:hei} we report total (narrow+outflow+BLR) line fluxes of the seven \hei\ lines detected in the integrated spectrum of GS\_3073 (see Figure~\ref{f:3x3spec}), normalised to the total line flux of \hei$\lambda4471$. We have verified that choosing a larger aperture would not significantly modify the line ratios. We note that \hei$\lambda4713$ is uncertain due to blending with [Ar\,{\sc{iv}}]$\lambda4711$. \hei$\lambda7065$ is particularly strong compared to theoretical predictions \citep[e.g.][]{Smits96, Benjamin99, DelZanna22}, which is also seen in some other data sets \citep[see e.g.][]{Benjamin99}. Knowledge of relative \hei\ intensities may be useful in reproducing more complex spectra of type 1 AGN showing prominent Fe~{\sc{ii}} emission in addition to \hei\ \citep[see e.g.][]{VeronCetty04, Perna21}.

\begin{table}
\caption{Total (narrow+outflow+BLR) emission line fluxes of the \hei\ lines detected in the integrated spectrum of GS\_3073 as constrained through our best fit (Section~\ref{s:fitting}), normalised to \hei$\lambda4471$.}
\begin{tabular}{lc}
\hline
\hline
    Line & Flux/$F_{\rm He\,I\lambda4471}$ \\
\hline 
    \hei$\lambda4471$  & $1.00$ \\
    \hei$\lambda4713$$^a$ & $0.6\pm0.2$ \\
    \hei$\lambda4922$  & $0.9\pm0.4$ \\
    \hei$\lambda5876$  & $10.1\pm2.6$ \\
    \hei$\lambda6678$  & $5.4\pm1.4$ \\
    \hei$\lambda7065$  & $12.5\pm3.2$ \\
    \hei$\lambda7281$  & $1.3\pm0.5$ \\
\hline
    \multicolumn{2}{p{1.0\columnwidth}}{$^a$Blended with [Ar\,{\sc{iv}}]$\lambda4711$.}
\end{tabular} 
\label{t:hei}
\end{table}

\section{Environment of GS\_3073 at $\sim1\micron$}\label{a:r100_comp}

\begin{figure}
	\centering
	\includegraphics[width=0.5\columnwidth]{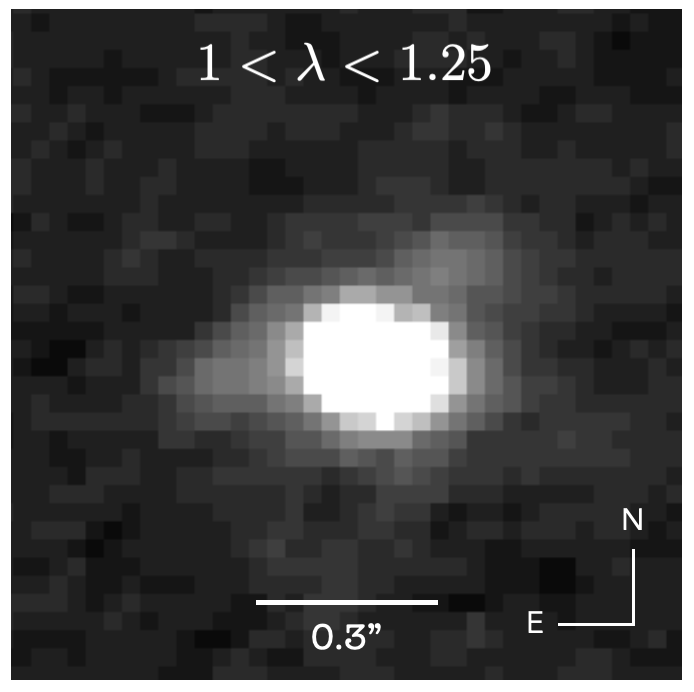}
	\caption{Image from the R100 data cube combined with a pixel scale of $0.03\arcsec$ in the wavelength range $1\micron<\lambda<1.25\micron$. North is up and East is to the left. In addition to the bright central source, faint emission is detected in the East and North-West, either associated with the AGN host galaxy GS\_3073, or low-mass companions.}
	\label{f:companions}
\end{figure}

In Figure~\ref{f:companions} we show a log-scale map based on the R100 data, summing the flux in the observed wavelength range $1\micron<\lambda<1.25\micron$. In addition to the central core of GS\_3073, two regions of faint emission become apparent in the East and North-West. 
The faint flux to the East could arguably be associated with a low-mass companion, coincident with a kinematically distinct region of high narrow line velocities (see top centre and middle centre panels in Figure~\ref{f:kinmaps}). 
To the North-West there is no distinct kinematic feature visible in our spatially-resolved maps, but the \oiii\ outflow velocities are more redshifted in this region. It is conceivable that we are actually picking up emission from a faint companion.
We note that \cite{Grazian20} speculate that the [C\,{\sc{ii}}] emission associated with GS\_3073 \citep{LeFevre20, Barchiesi22} could indicate an ongoing merger with a dusty companion, possibly fuelling the AGN activity of GS\_3073.

\section{Integrated PRISM spectrum}\label{a:r100_spec}

\begin{figure*}
	\centering
	\includegraphics[width=\textwidth]{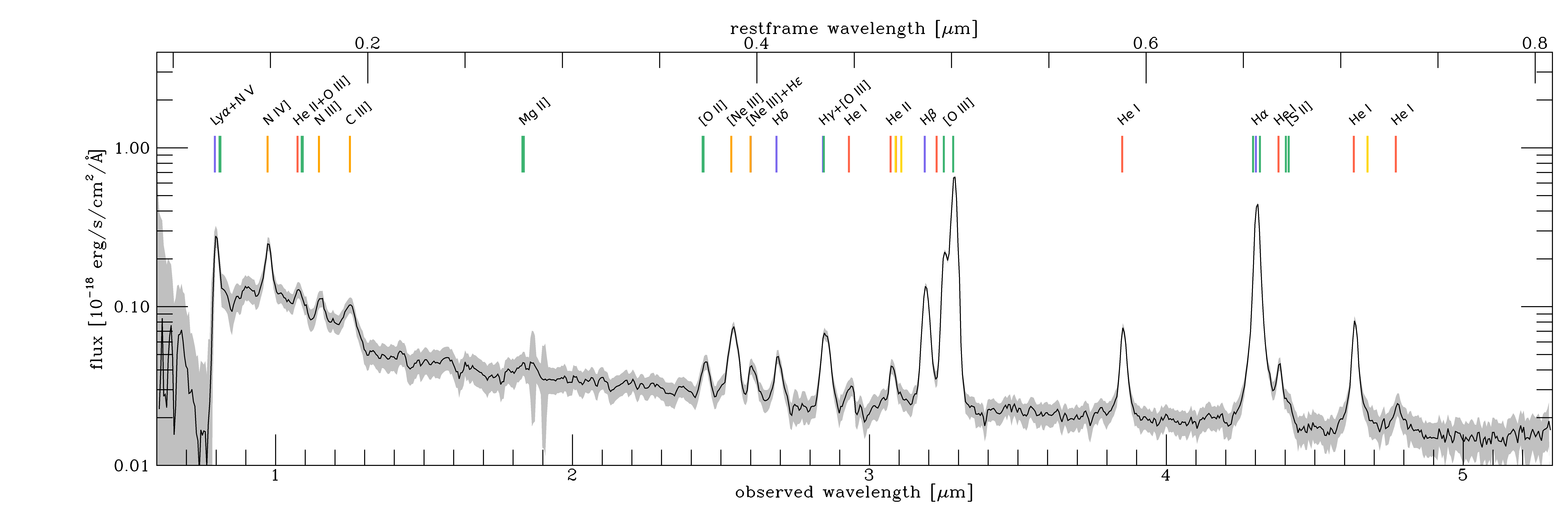}
	\caption{Integrated spectrum of the PRISM data extracted from a circular aperture containing the central 24 spaxels, flux-matched to the integrated G395H spectrum discussed in the main text, with flux in log scale. We indicate the positions of another 14 emission lines (or emission line doublets), in addition to what is discussed in the main text based on the G395H spectrum.}
	\label{f:r100}
\end{figure*}

In Figure~\ref{f:r100} we show the PRISM spectrum integrated over the central 24 spaxels, flux-matched to the integrated G395H spectrum discussed in the main text, with flux in log scale. In addition to the emission lines detected in the G395H spectrum and discussed in the main text, we indicate the positions of another 14 emission lines (or emission line doublets). These additional lines cover the range from Ly$\alpha$ to H$\gamma$.

%%%%%%%%%%%%%%%%%%%%%%%%%%%%%%%%%%%%%%%%%%%%%%%%%%

\label{lastpage}
\end{document}